\documentclass[twocolumn]{aastex62}
\usepackage{graphicx}
\usepackage{subfigure}
\usepackage{epstopdf}
\usepackage{amsmath} 
\usepackage{amsmath}
\usepackage{bm}
\usepackage{graphicx}
\usepackage{multirow}
\usepackage{enumerate}
\usepackage{amsmath}
\usepackage{bm}
\usepackage{graphicx}
\usepackage{natbib}
\usepackage{color}
\hypersetup{colorlinks,
linkcolor=blue,
citecolor=blue,
filecolor=blue,
urlcolor=blue}

\usepackage{rotating}

\usepackage{soul}

\usepackage{enumerate}
\usepackage{textcomp}

\usepackage{multirow}
\usepackage{enumerate}
\begin{document}
\title{Does modified gravity predict fast stellar bars in spiral galaxies?}
\correspondingauthor{Mahmood Roshan}
\email{mroshan@um.ac.ir}
\author{Neda Ghafourian}
\affiliation{Department of Physics, Faculty of Science, Ferdowsi University of Mashhad P.O. Box 1436, Mashhad, Iran}
\author{Mahmood Roshan}
\affiliation{Department of Physics, Faculty of Science, Ferdowsi University of Mashhad P.O. Box 1436, Mashhad, Iran}
\affiliation{School of Astronomy, Institute for Research in Fundamental Sciences (IPM), 19395-5531, Tehran, Iran}
\author{Shahram Abbassi}
\affiliation{Department of Physics, Faculty of Science, Ferdowsi University of Mashhad P.O. Box 1436, Mashhad, Iran}
\begin{abstract}
The evolution of disk galaxies in modified gravity is studied by using high-resolution N-body simulations. More specifically, we use the weak field limit of two modified gravity theories, i.e., nonlocal gravity (NLG)  and scalar-tensor-vector gravity known as MOG, and ignore the existence of dark matter halo. On the other hand, we construct the same models in the standard dark matter model and compare their dynamics with the galactic models in modified gravity. It turns out that there are serious differences between galactic models in these different viewpoints. For example, we explicitly show that the galactic models in modified gravity, host faster bars compared to the dark matter case. On the other hand, final stellar bars are weaker in modified gravity. These facts are not new and have already been reported in our previous simulations for exponential galactic models. Therefore, our main purpose in this study is to show that the above-mentioned differences, with emphasis on the speed of the bars, are independent of the initial density profile of the adopted disk/halo. To do so, we employ different profiles for the disk and halo and show that the results remain qualitatively independent of the initial galactic models. Moreover, a more accurate method has been used to quantify the kinematic properties of the stellar bar. Our results imply that contrary to the dark matter models, bars in modified gravity are fast rotators which never leave the fast-bar region until the end of the simulation.
\end{abstract}


\section{Introduction}
It is well understood that the dynamics of spiral galaxies is substantially influenced by the existence of dark matter particles. More specifically, the dark matter halo is the main mass component of each galaxy and plays a central role in many secular phenomena in the disk. For example, there is a direct correlation between the stellar bar evolution and the dark matter halo properties. It is well established in the literature that a rigid halo can suppress the bar instability and prevent the formation of strong bars \citep{1973ApJ...186..467O}. On the other hand, a responsive halo does not necessarily suppress the instability. The angular momentum transfer between the halo and the disk can expedite the bar formation \citep{2002ApJ...569L..83A}. The pattern speed of the bar is another important observable which is directly related to the properties of the surrounding halo. Because of the dynamical friction between the halo particles and the bar, the pattern speed decreases with time. This behavior is seen in many galactic numerical simulations, for example, see \cite{2000ApJ...543..704D}. As expected, the cosmological simulations also confirm the pattern speed slow-down \citep{2017MNRAS.469.1054A}. Therefore we may conclude that simulations in particle dark matter paradigm predict slow bars.

On the other hand, it is well-known in the literature that almost all the observed pattern speeds are fast \citep{2011MSAIS..18...23C,2020MNRAS.491.3655G}. Furthermore, it seems that bars are formed fast and remain as fast rotators during the cosmological time scale \citep{2012A&A...540A.103P}.This fact is considered as a challenge for the standard cosmological model. Of course, there are some proposals to address this problem. For example, it has been shown in \citep{2017PhRvD..95d3541H} that the ultralight fuzzy dark matter particles lead to a substantially less dynamical friction. Consequently, this kind of dark matter particles does not decrease the bar pattern speed via dynamical friction.

Based on the fact that dark matter particles have been not yet detected, there is another approach, i.e., modified gravity, to the above-mentioned challenge. Modified gravity theories aiming at replacing dark matter particles with viable and self-consistent modifications to general relativity (GR) have a long story \citep{2012PhR...513....1C}. In this paper, we will discuss two theories:  nonlocal gravity (NLG) proposed by \cite{2009PhLB..673..279H} to include nonlocal features of gravitation, and the scalar-tensor-vector theory of gravity known as MOG proposed to address the dark matter problem \citep{2006JCAP...03..004M}. Besides Modified-Newtonian Dynamics (MOND) \citep{1983ApJ...270..365M}, these theories are the main alternative models to dark matter in the literature. One may note that these theories are different in nature and phenomenology from the other branch of modified theories of gravity which try to address the dark energy enigma in cosmology.  

Although NLG and MOG are based on completely different postulates and motivations, their weak field limit is similar. Both theories lead to Yukawa like corrections to the gravitational force between point masses. These corrections help to explain the flat rotation curves of spiral galaxies as well as the mass discrepancy in galaxy clusters, for more details see \cite{2014PhRvD..89j4011R}; \cite{2013MNRAS.436.1439M}; \cite{2014MNRAS.441.3724M}. The formation and evolution of stellar bars in these theories have been investigated in \cite{2019ApJ...872....6R} and \cite{2018ApJ...854...38R} using high-resolution simulations. For low-resolution but comprehensive sets of simulations in MOG, we refer the reader to \cite{2017MNRAS.468.4450G}. Since there is no dark matter halo in galactic models in modified gravity, there is not significant dynamical friction in the system. Consequently, it is shown in  \cite{2019ApJ...872....6R} and \cite{2018ApJ...854...38R} that bars are faster compared to the standard case and the pattern speed is almost constant with respect to time. This fact appears as a serious difference between modified gravity and particle dark matter. Combined with the relevant observations, this deviation between these different viewpoints may help to distinguish between particle dark matter and modified gravity in galactic scales. However, simulations presented in  \cite{2019ApJ...872....6R} and \cite{2018ApJ...854...38R}, have three main restrictions and it is necessary to resolve them to ensure that modified gravity, unlike to the standard picture, predicts the existence of fast bars in spiral galaxies. 

Let us briefly mention the three restrictions: i) fast bars appear in simulations of modified gravity with exponential disks. This does not necessarily mean that bars are fast in other density profiles. In other words, it is necessary to show that the main results of our previous simulations are model-independent. ii) there is no bulge component in the simulations. For more realistic galactic models it is necessary to include the bulge component. iii) we know that gas plays a crucial role in the secular evolution of galactic disks. Therefore, hydrodynamic simulations are required in order to strengthen and confirm the results obtained from N-body simulations.

Our main motivation in this paper is to investigate the first restriction by constructing different galactic models. In other words, we show that our previous results are model-independent. As mentioned, this is a crucial test in order to ensure the significant deviations between modified gravity and particle dark matter. To do so, we first briefly review the main features of NLG and MOG in section \ref{nlgmog}. Then we discuss the numeric methods, the initial conditions and, in particular, the GALAXY code in section \ref{methods}. Furthermore, we present the results of a series of N-body simulations in section \ref{results}. Finally, we end up with the discussion and conclusion section \ref{conc}.

\section{Yukawa corrections: manifestation of nonlocality or coupled Proca vector field?}\label{nlgmog}

As we already mentioned, NLG and MOG induce Yukawa corrections to the gravitational force between point masses. Note that to investigate the dynamics of a spiral galaxy that lies in the Newtonian regime, we need the weak field limit of the gravitational theories. This point combined with the fact that there is no screening behavior in NLG and MOG causes a substantial simplicity in our analysis. In other words, one may conveniently linearize the field equations of the theories and find the modified version of the Poisson equation. Then it is straightforward to find the gravitational force between point particles. The absence of screening behavior allows us to use the superposition principle to find the gravitational field of an extended mass distribution. This kind of linearization is not allowed in $f(R)$ gravity theory where the inherent scalar degree of freedom of the theory possesses screening mechanism \citep{2007PhRvD..76f4004H}. 

The weak field limit of NLG and MOG has been reviewed in \cite{2019ApJ...872....6R} and \cite{2018ApJ...854...38R} respectively. Also for more details one may see \cite{2014PhRvD..89j4011R} and \cite{2014PhRvD..90d4010R}. Therefore we do not repeat the details here. However, it is important to mention that both theories introduce the following corrections to the Newtonian gravitational force between point particles
\begin{equation}
\mathbf{f}=-\frac{m_1 m_2 G}{r^2}\Big(1+\alpha-\alpha(1+\beta r)e^{-\mu r}\Big)\hat{\mathbf{r}}
\end{equation}
where $\alpha$ and $\mu$ are constant parameters, and $\beta=\mu/2$ in NLG and $\beta=\mu$ in MOG. These parameters have been fixed by fitting to rotation curve data of spiral galaxies. It is interesting to mention that although both theories lead to the same functionality for the gravitational force, the origin of the Yukawa corrections is completely different. In MOG, besides the metric tensor, there are two extra scalar fields and a Proca vector field which is coupled to matter. In this case, $\alpha$ and $\mu$ are related to the current background values of the scalar fields. Furthermore, $\mu$ plays the role of the mass of the vector field. These parameters, in principle, are functions of time.  Albeit \cite{2018JCAP...01..048J} using dynamical system approach for the cosmology of MOG has shown that $\mu$ does not vary significantly during the thermal history of MOG. However, for simplicity, we assume that they are constant. It is also necessary to mention that recently \cite{2019PDU....25..323G} has been postulated that $\alpha$ and $\mu$ should be decreasing functions of the mass of the galaxy to explain the radial acceleration relation (RAR) discovered by \cite{2016PhRvL.117t1101M}.

On the other hand, in NLG $\alpha$ and $\mu$ appear in the weak field limit as a manifestation of the nonlocality in gravitational interaction. Interestingly, nonlocal features of gravity can effectively appear as a source for strengthening the gravitational force \citep{2009PhLB..673..279H}. One may note that modified gravity models which try to address the dark matter problem,  should increase the gravitational force without postulating the existence of dark matter particles.

 We should emphasize that by NLG we mean a nonlocal gravity theory introduced by \cite{2009PhLB..673..279H}. This theory is directly based on the interesting similarity between Maxwell's equations for electrodynamics and those of general relativity when written in the teleparallel formulation. This similarity is extremely helpful to add nonlocal features to gravity in a similar way that already developed in electrodynamics, for a comprehensive discussion see \cite{mashhoon2017nonlocal}. There are different types of nonlocal gravity theories in the literature based on different viewpoints and motivations. For a pioneering work in this direction see \cite{2009JCAP...08..023D}. 

In NLG the free parameters $\alpha$ and $\mu$ also vary with time. However, for the sake of simplicity, we assume that these parameters are constant. Of course for a more careful analysis, it is required to take into account the time dependence of these parameters. Albeit implementation of this requirement in the N-body simulations is not a simple task from technical point of view. On the other hand, NLG is not an action-based theory and in practice when dealing with real physical systems some substantial mathematical and conceptual complexities appear in the analysis. For a comprehensive review of NLG, we refer the reader to the book \cite{mashhoon2017nonlocal}. On the other hand, for recent developments in the theoretical aspects of the theory see \cite{Puetzfeld:2019wwo} and \cite{2020arXiv200111073P}.

\section{Numerical Method}\label{methods}
To study the evolution of the disks under the effect of dark matter (DM) halo / modified gravity, we use GALAXY, a high-resolution N-body code \citep{2014arXiv1406.6606S} which is ideal for investigating collisionless stellar systems \citep{2019ascl.soft04002S}. The code constructs initial conditions for the systems in equilibrium and then evolves it according to time, by using a leapfrog method. The models under study are flat disks without any central bulge. As we mentioned earlier, simulations in modified gravity do not employ a DM halo. So in the absence of the bulge, the models in modified gravity are considered to be one-component systems. However, for the Newtonian model, we use a spherical live DM halo, whose properties will be described in the next section. In the DM model, we use a hybrid mesh consisting of a spherical three-dimensional (S3D) mesh plus a cylindrical polar three-dimensional (P3D) mesh, for the DM halo and the baryonic disk, respectively. While in the single-component models of modified gravity, MOG and NLG, only a P3D mesh is used.

\subsection{Initial Conditions}
As mentioned earlier, the main purpose of this work is to investigate the model-independency of previously reported results on the distinction of the disk evolution in modified gravity and dark matter. Since the previous studies have focused on Exponential (EXP) disks and Plummer halos \citep{2019ApJ...872....6R,2018ApJ...854...38R}, to assure the independence of the results to the adopted disk/halo models, we first changed the disk's density distribution from EXP to Kuzmin-Tommre (KT) while we keep the live Plummer halo model unchanged (LPH model). In the next step, we change both the halo and disk profiles. In this case, we employ a  KT disk and a live isothermal halo (LIH model). To investigate further, we also consider another halo model of Hernquist (LHH model). This halo model has a cusp in the inner radii, which increases the initial rotational velocity in the inner radii and influences the evolution of the disk. As we will discuss comprehensively, the dynamical friction plays a central role in discriminating between dark matter and modified gravity. On the other hand, it is natural to expect that this mechanism works differently in the cored and cuspy halo models. Therefore, to draw a complete and consistent picture it seems necessary to compare both the cored and cuspy types with modified gravity models. For this purpose we include the LHH model in the simulations.
 
In this study, the KT disk, model 1 of \citep{1963ApJ...138..385T}, has a surface density of the form
\begin{equation}
\Sigma(R)=\frac{M_d}{2 \pi R_d^2}\left[\left(\frac{R}{R_d}\right)^2+1\right]^{-3/2},
\end{equation}
where $R$ is the radial coordinate in the cylindrical coordinate system $(R,\phi,z)$. On the other hand, the Plummer halo, a polytrope model of index 5, has the density
\begin{equation}
\rho(r)=\frac{3 M_h}{4 \pi b^3}\left[ 1+\left(\frac{r}{b}\right)^2  \right]^{-5/2}
\end{equation}
while for the simple cored Isothermal halo model \citep{1993MNRAS.260..191E}, the density is given by
\begin{equation}
\rho(r)=\frac{V_h^2}{4\pi G a^2}\frac{3+x^2}{(1+x^2)^2}.
\end{equation}
The Hernquist halo model \citep{hernquist1990analytical} also has density of the form
\begin{equation}
\rho(r)=\frac{M_h c}{2\pi r (r+c)^3}
\end{equation}
In the above equations, $M_d$ and $M_h$ are the disk and halo mass, respectively. $R_d$ is the radial disk scale length, $a$, $b$ and $c$ are characteristic radii of the halos, and $x=r/a$. Furthermore, $V_h$ in the Isothermal sphere is $V_h=\sqrt{GM_h/a}$. It should also be mentioned that the system of units is considered so that $M_d=R_d=G=1$, where $G$ is the gravitational constant. Therefore, the unit of velocity and time can be written as $V_0=(GM_d/R_d)^{1/2}$ and $\tau_0=R_d/V_0=(R_d^3/GM)^{1/2}$. As a suitable example, selecting $R_d=2.6$ kpc and $\tau_0= 10$ M yr, yields $M_d\simeq 4 \times 10^{10} M_{\odot}$ and $V_0\simeq 254$ km s$^{-1}$.
\begin{figure} 
 {\includegraphics[width=8cm]{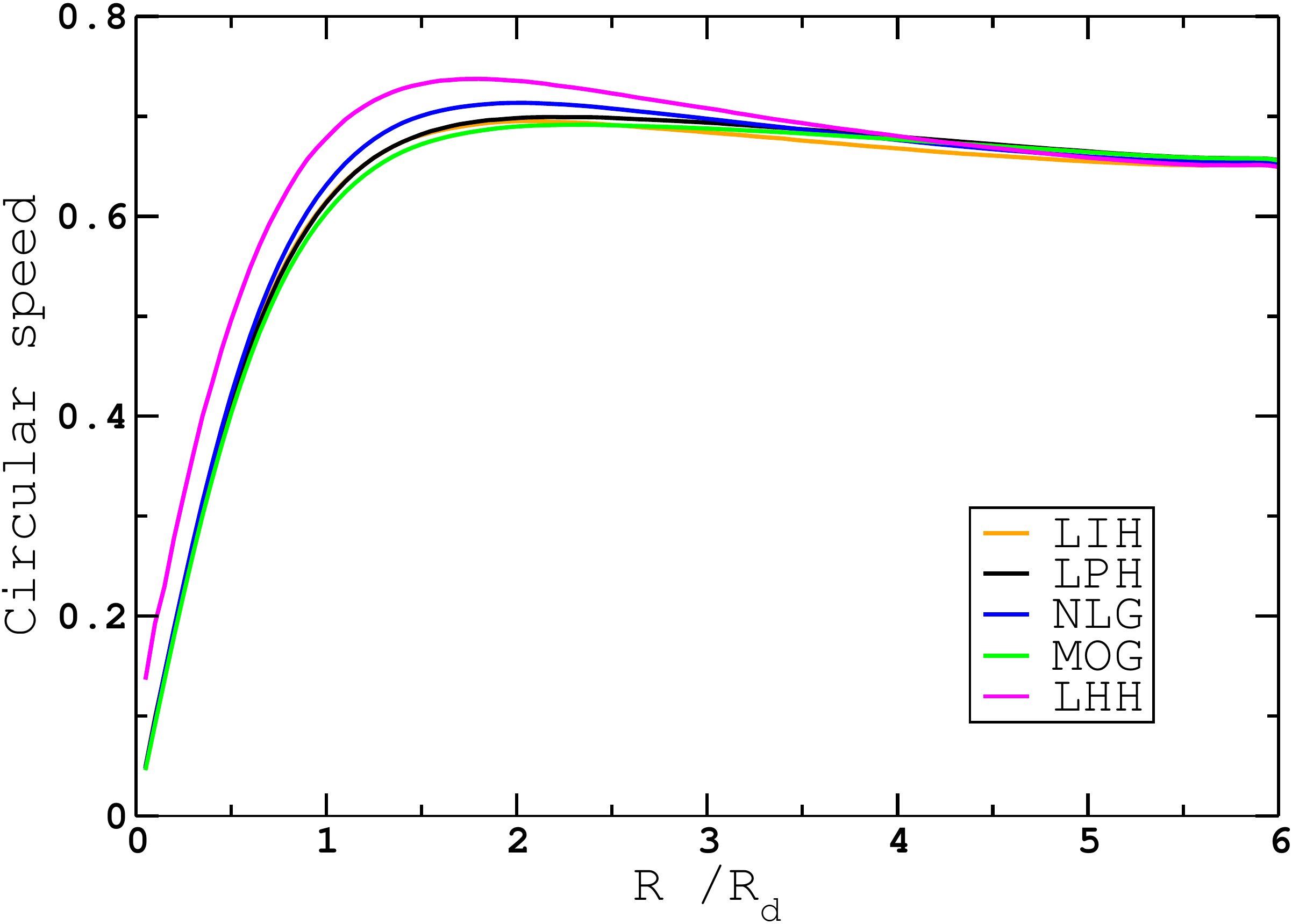}}
\caption{Initial rotational velocities for NLG (blue), MOG (green), LPH (black) and LIH (orange) models. The disk has a KT density distribution in all four models.}
\label{vt0}
\end{figure}

To make a reasonable comparison, we constrain our models to start their evolution from a similar state. Therefore, the models are constructed such that the initial rotational velocity of the four models is as close as possible. Furthermore, The initial velocity dispersion of the models has a very similar trend.  To do this, the employed free parameters of the DM halo density/modified gravity should be varied until achieving the same initial state for the baryonic matter in all the models. Notice that we use the same parameters for the disk density in all the models.

The truncation radius of the KT disk is chosen to be at $r=6R_d$, where the density starts to taper smoothly from $r=5R_d$ by using a cubic function. For our Plummer model, the halo mass is chosen as $M_h=3.6 M_d$ and the halo scale length as $b=8.5 R_d$, where the truncation radius is at $r= 3.4 b$.  In the Isothermal model, the halo mass is $M_h=1.7M_d$ and its scale length is $a=6R_d$ with truncation radius at $r=10a$. In the Hernquist model, the adopted halo mass is $M_h=14 M_d$, the scale length is $c=19 R_d$ and the truncation radius is at $r=8c$. As mentioned earlier, the cusp in the Hernquist halo results in a relatively higher rotational velocity in the inner radii, however, with this choice of parameters, the rotation curve in the outer radii matches the other models very well. The LHH model seems too massive compared to the other models. However, regarding the Gauss's theorem, at least at the beginning of the simulation, only the dark matter mass inside a sphere with radius $6 R_d$, i.e. $M(6R_d)$, contributes to the gravitational force. It is easy to show that $M(6R_d)\simeq 0.8$, $0.7$, $0.85$ for LHH, LPH and LIH models respectively. In other words, in all the models the dark matter mass fraction inside $r=6R_d$ is around $0.4$. Therefore one may ensure that all the models start with suitable initial conditions.

 For the NLG model, selected value for the free parameters are $\alpha=10.9$ and $\mu=0.0045 R_d^{-1}$. Also, for the MOG model we employ $\alpha=2.8$ and $\mu=0.031 R_d^{-1}$. The resulted rotational velocity of our models is illustrated in Fig. \ref{vt0}.  

For all the models, the Toomre parameter Q \citep{1964ApJ...139.1217T} is considered to be 1.5. This value is high enough for our models to avoid the disk from local instability and fragmentation.

It should also be mentioned that for every active component (DM halo/disk) in our main simulations, we have adopted $N=2\times 10^6$ particles. However, to make sure that the results are independent of the particle number, we have performed the simulations for a higher number of particles (Section \ref{tests}).

It should be recalled that the adopted mesh for the disk and DM halo is a cylindrical polar (P3D) and a spherical (S3D), respectively. The grid size selected for our S3D mesh is $1001 \times 2501$ where the first number is the number of radial grid shells and the second is the radius of the grid outer boundary. For the P3D mesh, we have $193 \times 224 \times 125$, where the numbers demonstrate the number of mesh points in radial, azimuthal and vertical direction, respectively. Also, "lscale", which is the number of mesh points on the length unit, is chosen to be 12.5. Furthermore, the softening length in the P3D mesh is $\epsilon=0.16 R_d$. This choice of mesh points is enough to certify that the majority of particles do not leave the mesh during the simulations. This property is checked for all the models and less than $2\%$ of the particles escape the mesh until the end of the simulation. 

It is necessary to mention that the computation of the potential is achieved by using sectoral harmonics $0\leqslant m \leqslant 8$ in the P3D mesh and surface harmonics $0\leqslant l \leqslant 8$ in the S3D. More details could be found in the \href{http://www.physics.rutgers.edu/~sellwood/manual.pdf}{online manual}.

The duration of our simulations is selected to be $\tau=800\tau_0$ and the time step is $\Delta \tau=0.01 \tau_0$. It should also be mentioned that three different time zones with factors of $1$, $2$ and $4$ $\Delta \tau$  are selected to account for the evolution of particles in regions of different density; Shorter time steps are required for particles in denser regions. However, to ensure that the results are not affected by the choice of the time step, we also checked the results for lower values of $\Delta\tau$ (Section \ref{tests}). Moreover, the grid is recentered every 16 steps to avoid numerical artifacts.

According to the selected options, the code computes the initial positions and velocities of the particles to form the initial equilibrium state of the system. It should be noted that if the model consists of more than one component, it is necessary to find the distribution function (DF) in the composite system and then choose the particles from the resulted DF. Accounting for this matter in our DM model, we used the Compress algorithm \citep{1980ApJ...242.1232Y,2005ApJ...634...70S} implemented in the code, which finds the potential and changes the halo density so that the equilibrium condition in the presence of the disk is met. The initial positions and velocities of the system are then obtained. Now, let us discuss the results in the following section.

\section{Results}\label{results}
To study the systems under consideration, we start from the initial conditions described in the previous section, and then the evolution of the system is monitored with time. To have a better understanding of the effect of our modified gravity theories in contrast to the DM halo, we study the following quantities. The growth of the bar, its pattern speed and power spectra, the evolution of the Toomre parameter which is a measure of disk's heating, the vertical thickness of the disk and finally the $\mathcal{R}$ parameter. This parameter quantifies the pattern speed of the bars. There are explicit observations for $\mathcal{R}$ which could be useful in contrasting different gravitational models. The comparison between our models, regarding these quantities, will be discussed in the following subsections. As it will be demonstrated, the behavior of the disk under the influence of dark matter is distinctive in comparison to modified gravity.
\begin{figure} 
{\includegraphics[width=0.45\textwidth]{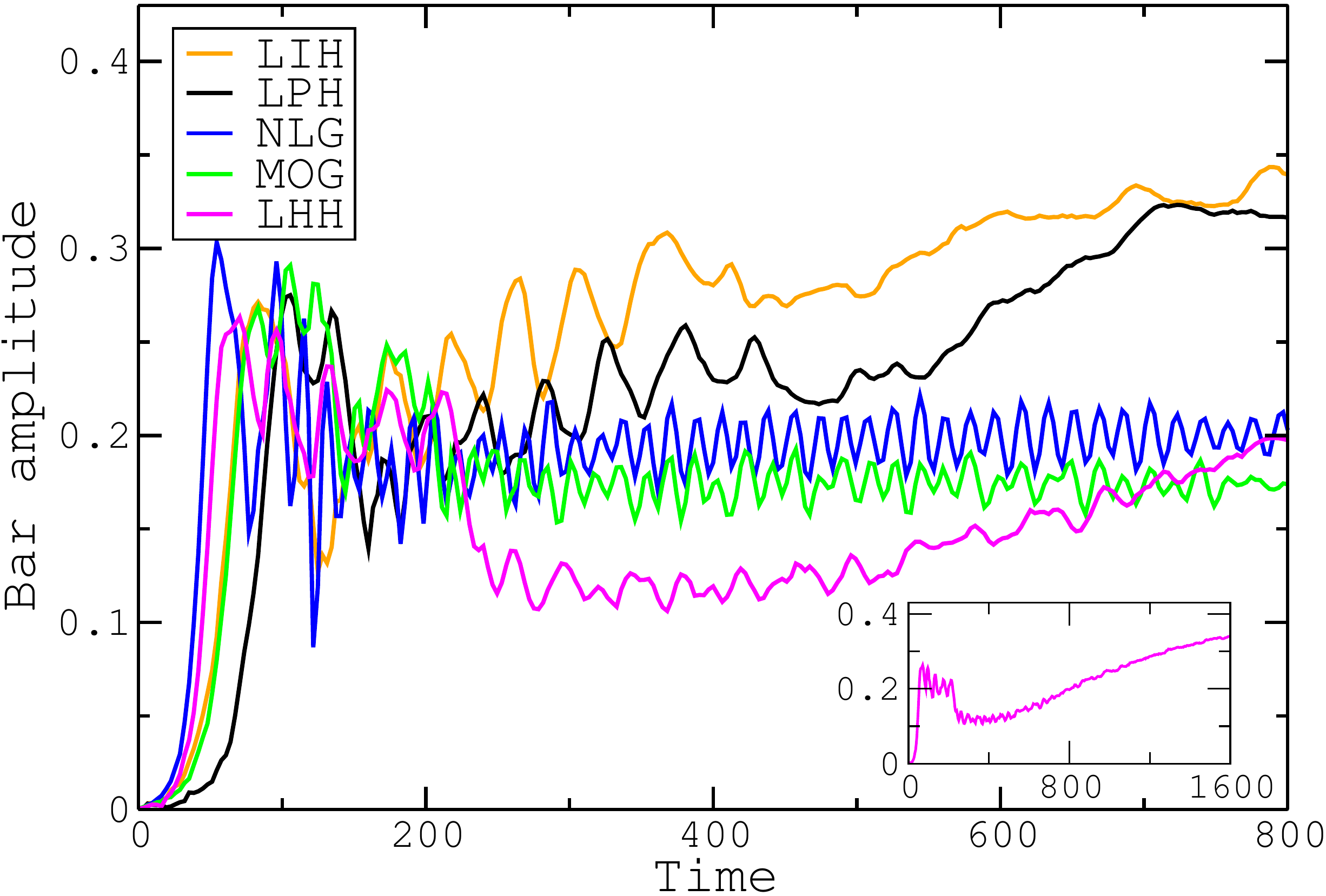}}
\caption{Time evolution of bar amplitude for our models. Bar growth starts faster in the NLG model (blue) and then the amplitude decreases and oscillates about an almost constant value. However, in the LPH and LIH Halo models (black and orange), the growth rate is lower, but the bar amplitude grows to higher values. The behavior of the MOG model (green) is similar to NLG but with a lower growth rate. The time evolution of bar amplitude for the LHH halo model is slower in comparison to the other DM models, however, the sub-plot shows that it reaches to the same value at later times.}
\label{bar}
\end{figure}
\subsection{Bar Growth}\label{s.bar}
In order to investigate the formation and evolution of the bar in our models, we use the Fourier expansion of the mass distribution of the particles in the disk plane. It is straightforward to show that $m$th Fourier coefficient is written as
\begin{equation}\label{eq-bar}
A_m(\tau)=\left| \sum_j \mu_j e^{i m \phi_j}\right|,
\end{equation} 
where $\mu_j$ is the mass and $\phi_j$ is the cylindrical polar angle of the $j$th particle at time $\tau$. It should be noted that we have used the same mass for all of the particles in our simulation, so every particle has an equal contribution to the total mass of the system. Also, we recall that the above summation is only over the disk's particles and naturally does not take the dark matter particles into account. Considering the above equation, one could find the bar/spiral amplitude by calculating the ratio of $A_2/A_0$. Although $m=2$ includes both the spiral and bar-like features, since the spirals are short-lived patterns, it would be reasonable to use this quantity to account for the effect of the bar mode.

Fig. \ref{bar} shows the evolution of the bar for our models. From this figure, it is apparent that the bar amplitude in both modified gravity theories starts with an initial peak, then after a drop, it continues to oscillate around an almost constant value. Compared to MOG, the formation of the bar is faster in NLG and the bar is stronger. Also, the bar amplitude reaches its constant value at $\tau\simeq 150$, while in the MOG model it happens later at about $\tau=200$. In comparison to NLG, the bar formation in DM models starts later and at a lower rate. It also has an initial peak which happens at $\tau \simeq 75$ for the Isothermal (orange) and at $\tau \simeq 100$ for the Plummer model, but after a drop at $\tau\simeq 170$, it starts to grow again with an almost constant rate. Both models experience a decreasing phase around $\tau \simeq 400$ and then it continues to grow until the end of the simulation.  However, it should be mentioned that although the bar strength in these two DM models have an overall similar trend and reach almost the same value at final states, the initial behavior in the formation and growth of the bar is not independent of the adopted halo profile. One can infer that, in comparison to the Plummer model, the bar starts to form in Isothermal model faster and at an earlier time which matches MOG in early stages. However, a clear discrepancy between the evolution of the disks in modified gravity and cored DM models are visible after about $\tau \simeq 200$, regardless of the selected mass model. 

On the other hand, studying the behavior of the LHH model, which includes a cuspy DM halo at the beginning of the simulation, shows some differences to the other DM models. From Fig. \ref{bar}, it is clear that the growth rate of the bar formation in the LHH model is higher. Moreover, the value of bar strength remains around its initial peak for a longer duration in comparison to the other halo models. Then, at about $\tau \simeq 250$, another phase in the evolution of the bar amplitude in the LHH model starts, in which the bar weakens. This period is also long-lasting in comparison to the other DM models. At about $ \tau \simeq 500$, the bar amplitude of the LHH model starts to grow again. However, at $\tau \simeq 800$, the bar in LHH is clearly weaker than LPH and LIH models. Since the evolution of the LHH model does not seem to be completed at this stage, we continued its evolution up to $ \tau =1600$ in the subplot of Fig. \ref{bar}. As one can see, the growing phase of the bar amplitude in this model continues to the end of the simulation. According to this figure, it could be concluded that although the details in the evolution of a cuspy DM model is different from a cored DM model and it shows some delays, its overall behavior does not change and is still very different from the models of modified gravity.

\begin{figure*}
\centerline{
   \includegraphics[width=18cm]{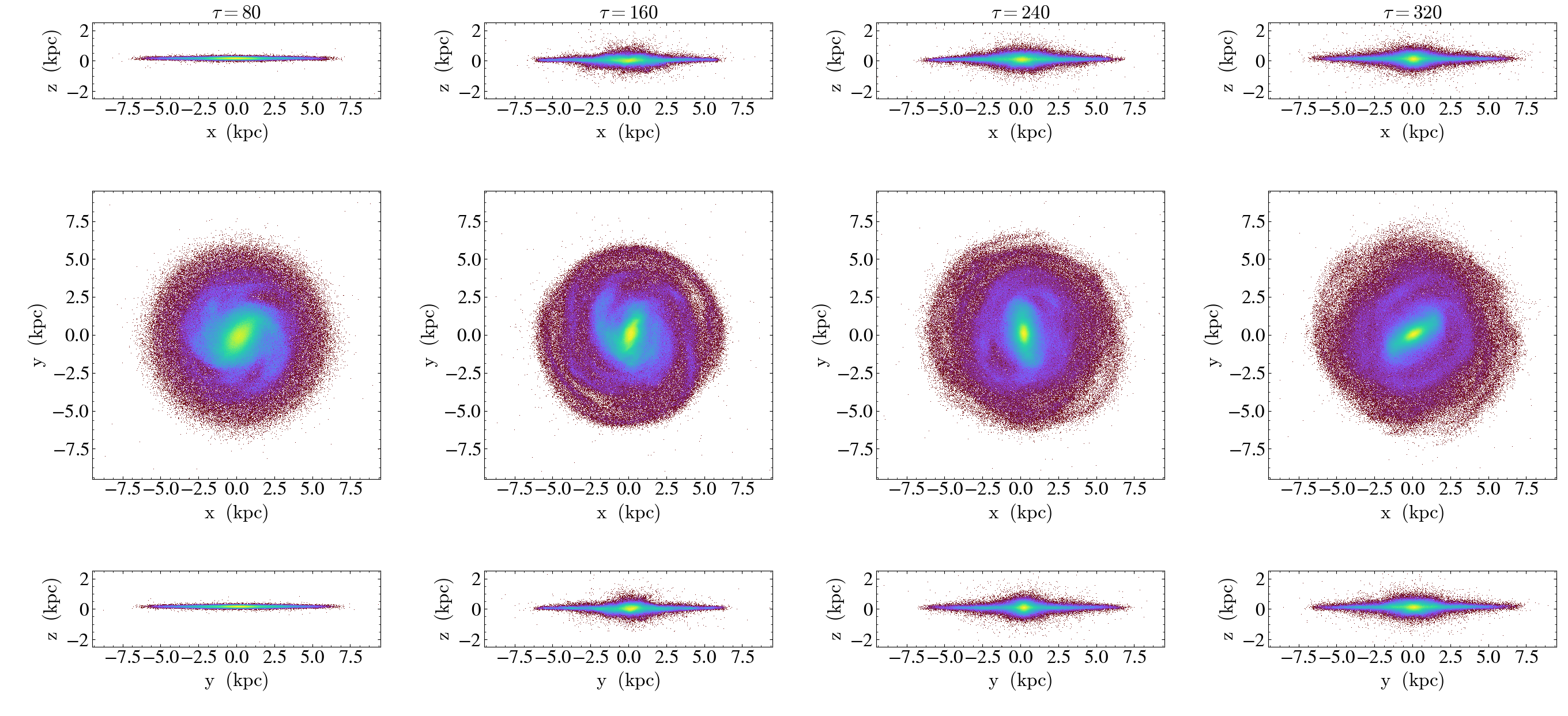}}\vspace{0.95 cm}
\centerline{
   \includegraphics[width=18cm]{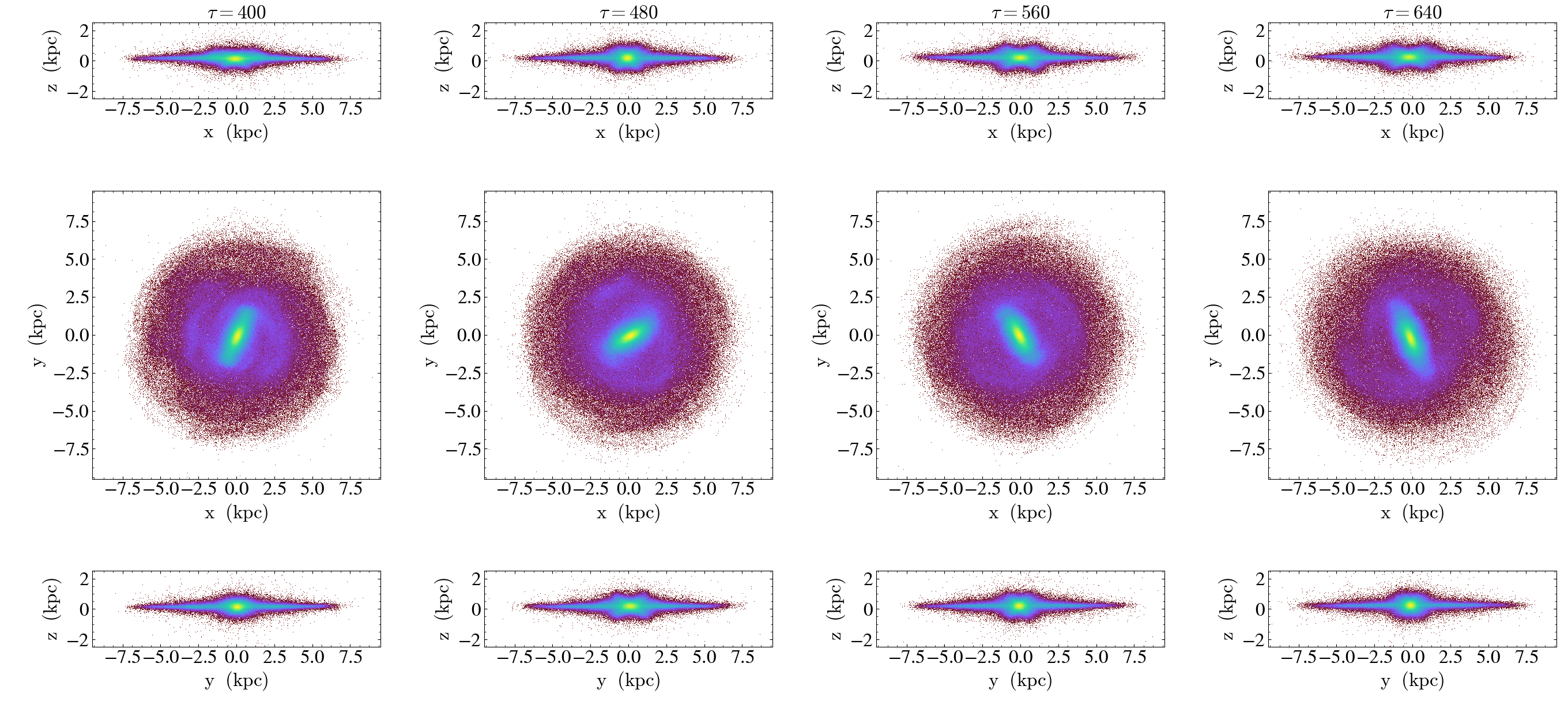}}
\caption{\small Projected positions of the particles in LPH disk. High-density regions are represented by lighter colors.  These plots are constructed using yt project \citep{2011ApJS..192....9T}.} 
\label{pos-LPH}
\end{figure*}

\begin{figure*}
\centerline{
   \includegraphics[width=18cm]{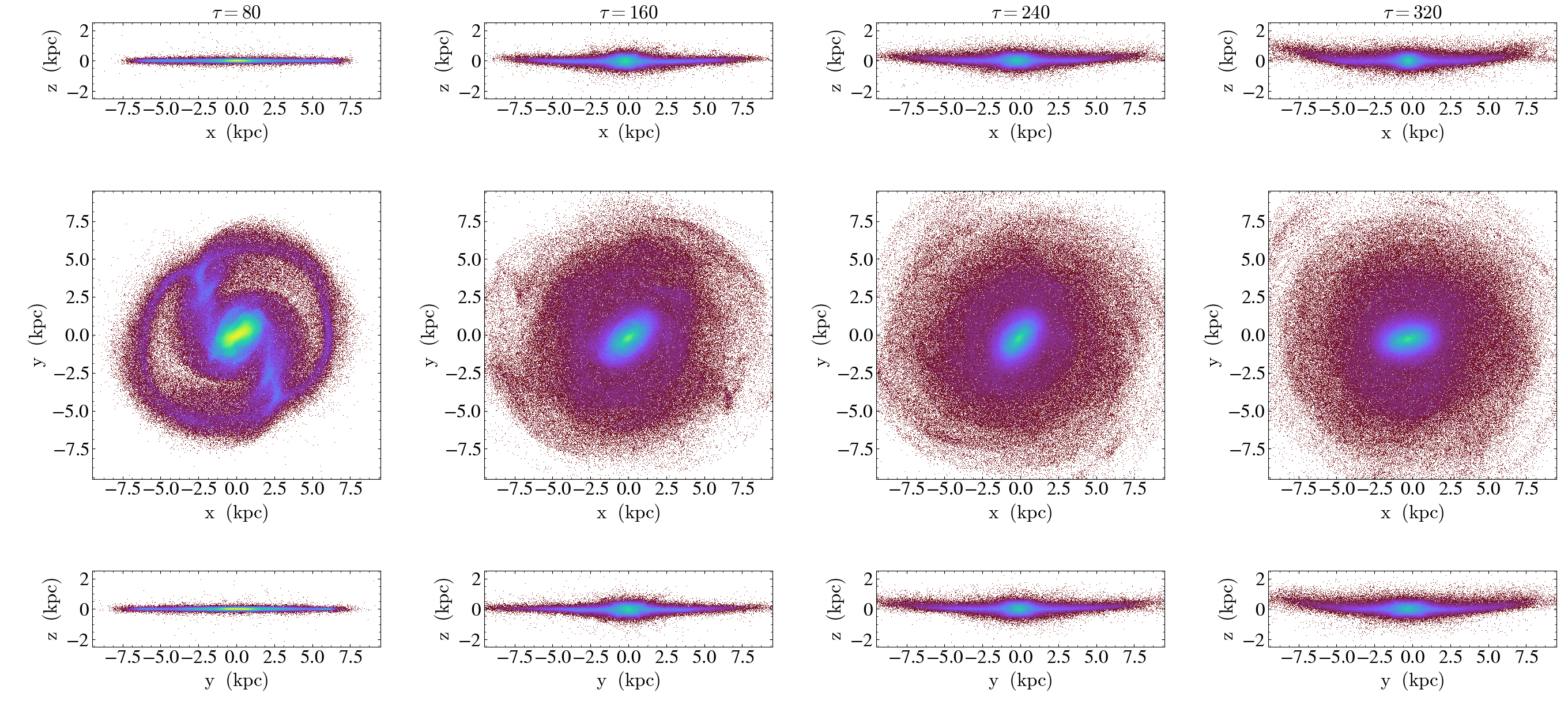}}\vspace{0.95 cm}
\centerline{
   \includegraphics[width=18cm]{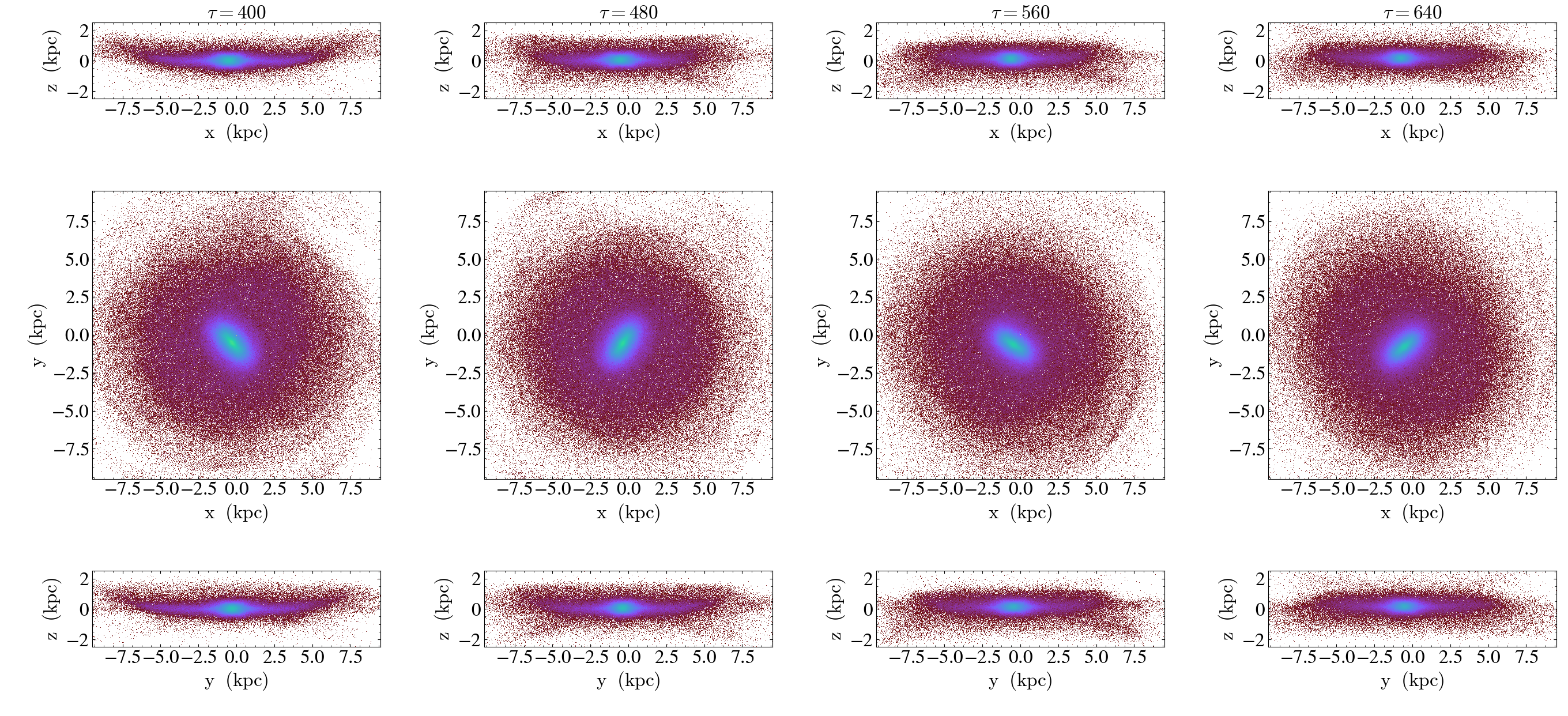}}
\caption{\small Projected positions of the particles in NLG disk.}
\label{pos_NLG}
\end{figure*}

\begin{figure*}
\centerline{
   \includegraphics[width=18cm]{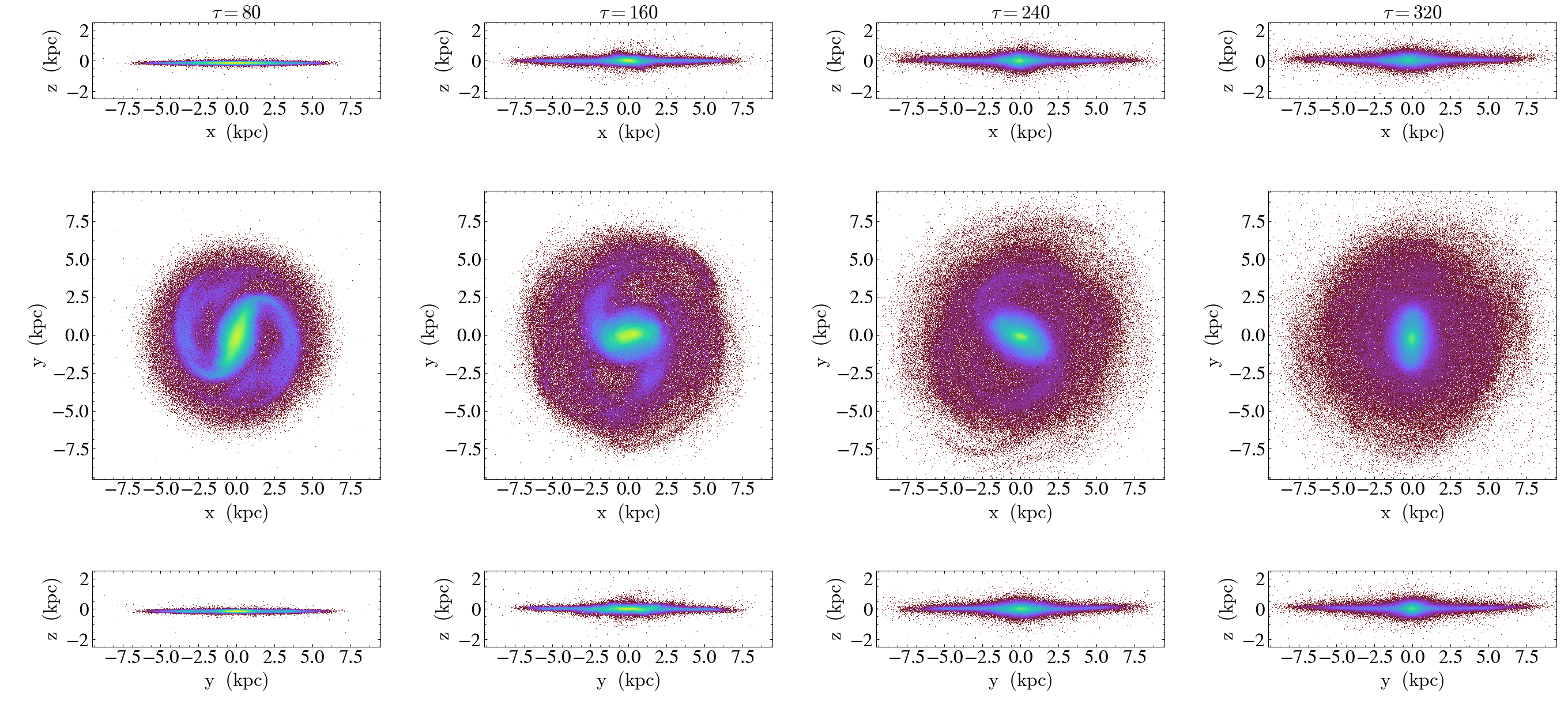}}\vspace{0.95 cm}
\centerline{
   \includegraphics[width=18cm]{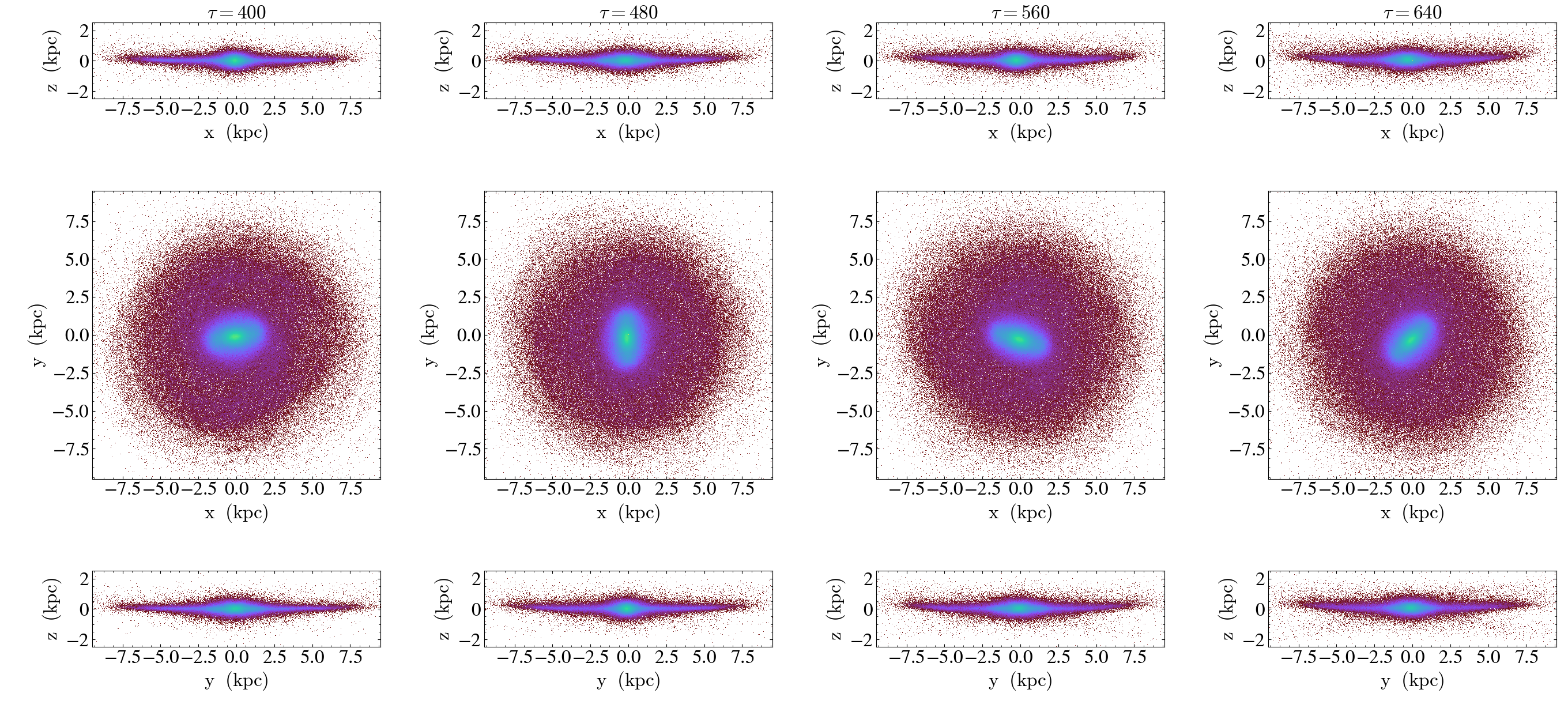}}
\caption{\small Projected positions of the particles in MOG disk.}
\label{pos_MOG}
\end{figure*}

\begin{figure*}
\centerline{
   \includegraphics[width=18cm]{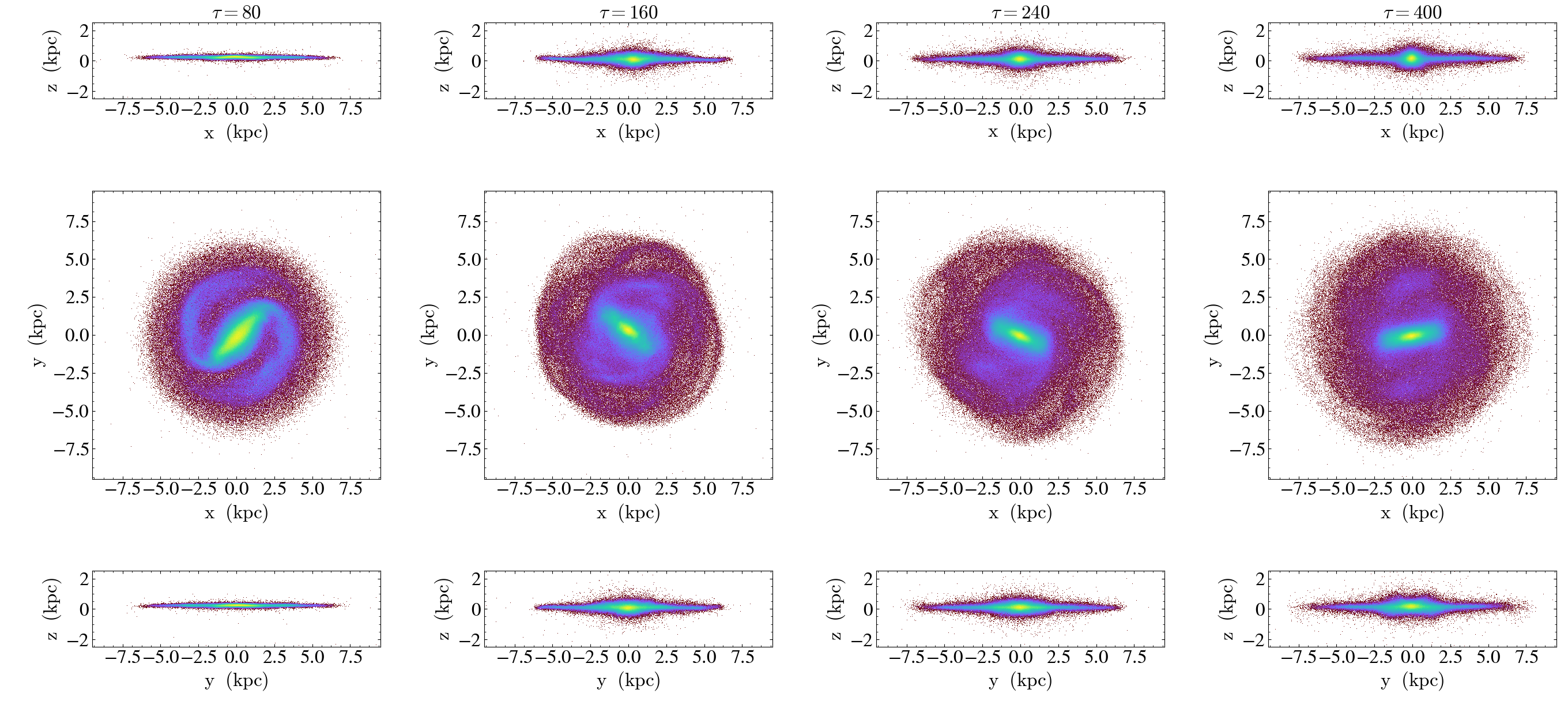}}\vspace{0.95 cm}
\centerline{
   \includegraphics[width=18cm]{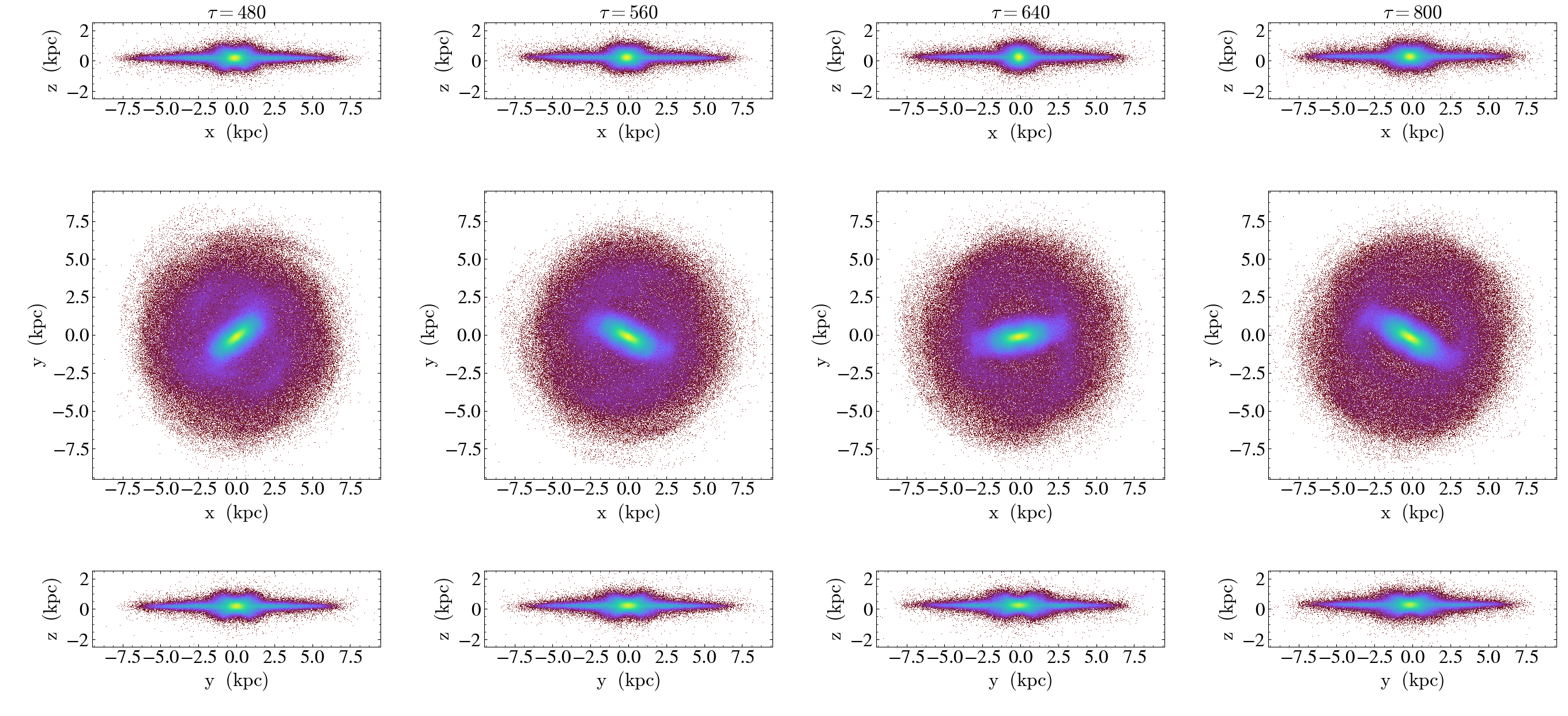}}
\caption{\small Projected positions of the particles in LIH disk.}
\label{pos_LIH}
\end{figure*}

\begin{figure*}
\centerline{
   \includegraphics[width=18cm]{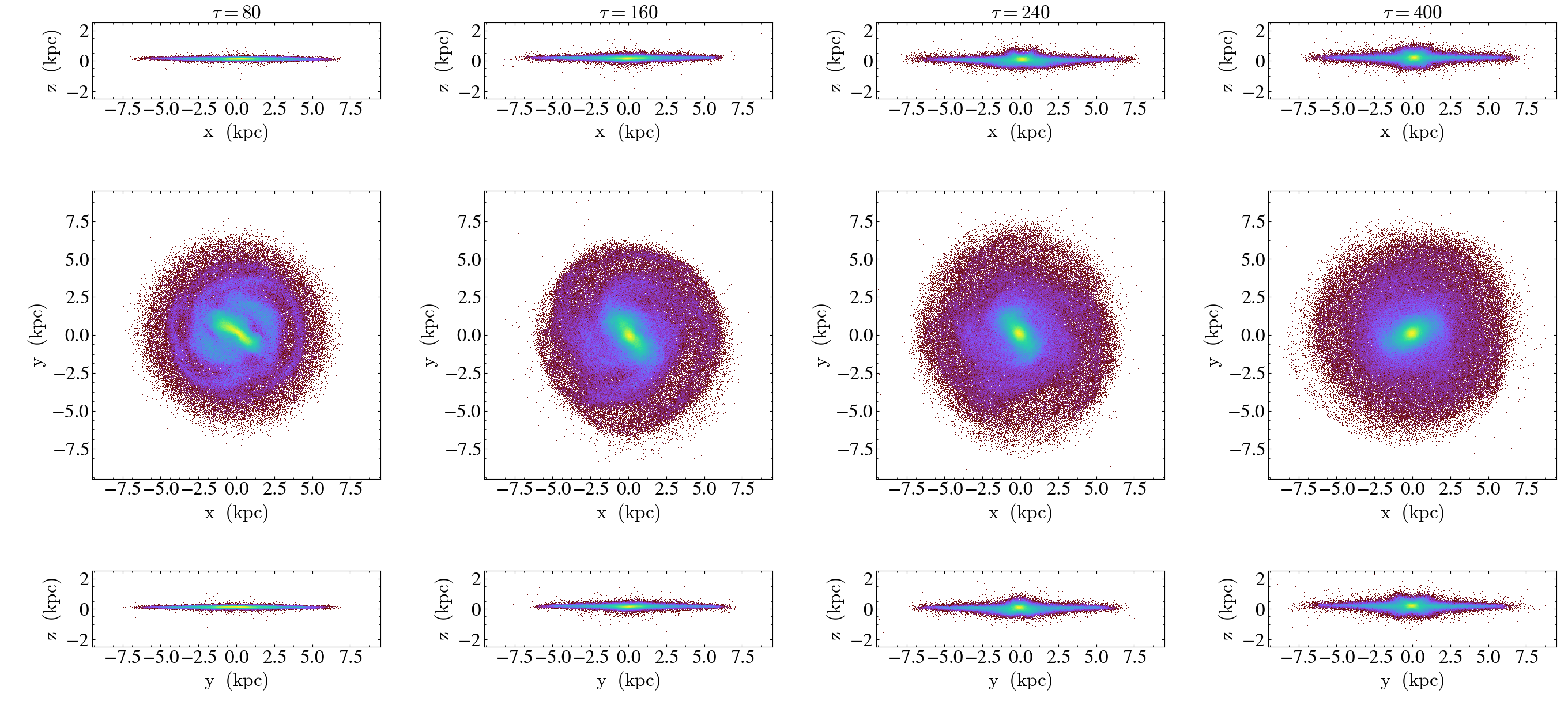}}\vspace{0.95 cm}
\centerline{
   \includegraphics[width=18cm]{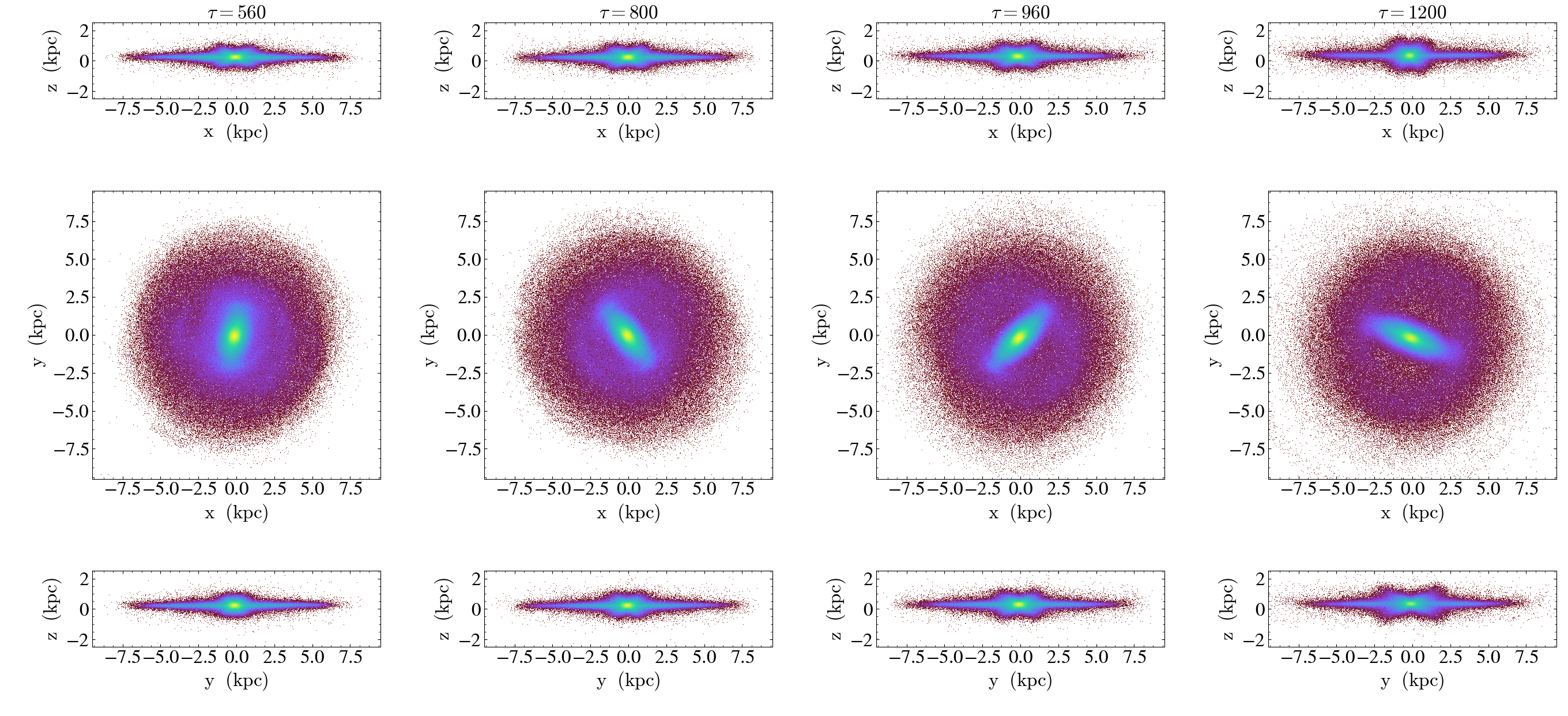}}
\caption{\small Projected positions of the particles in LHH disk.}
\label{pos_LHH}
\end{figure*}

According to this behavior, one may conclude that the comparison of bar evolution in our modified gravity/DM models, shows two different phases. In the first phase, which is common in all the five models, an initial strong non-axisymmetric $m=2$ mode forms in the disk which is short-lived and dissipates after a short period of time. As we mentioned, at least for the MOG case, it is a bit difficult to discriminate between modified gravity and particle dark matter in this phase.

On the other hand, the second phase, which starts at intermediate times $\tau\gtrsim 200$, reveals some significant differences between modified gravity/DM models. The bar continues with an almost constant value in modified gravity, while in the cored DM models it almost constantly grows until the end of the simulation. On the other hand, in the cuspy DM model it shows a period of constancy with a low value of bar amplitude and then, similar to the cored models, starts to grow until the end of the simulation. The other distinctive feature of bar amplitude in our models is the oscillations which are only present in modified gravity models. This comportment might be due to the presence of different beating modes that will be discussed in section \ref{spectrum}. However, one should be careful about the artifacts due to the choice of the center at which the quantities are being calculated. To be specific, let us assume that the center of the grid does not coincide with the centroid of the stellar bar. Then it is straightforward to verify that the bar amplitude given by $A_2/A_0$ would oscillate with roughly twice the frequency of the bar rotation ($\Omega_p$). However, as we will show in the next subsection, this is not the case and the oscillation period is not short enough to be half the rotational period of the bar. On the other hand, the GALAXY code uses McGlynn's method to find the position of the particle centroid \citep{1984ApJ...281...13M}. With a suitable choice of the method's single parameter, the method is more sensitive to the internal condensation of the particles, where the bar is formed and puts less weight on distant particles. Therefore the code appropriately finds the centroid of the bar. This means that the oscillations are not artifacts and reveal a real feature. Furthermore, as we already mentioned the grid is re-centered every 16 time-steps in order to minimize the numeric artifacts.

The same general behavior of the disks under modified gravity/DM halo was also presented in \citet{2019ApJ...872....6R}, where the same gravitational models are studied for different mass profiles. Therefore, it could be concluded that ignoring the first stage of the evolution, the significant differences between bar evolution in modified gravity and dark matter, remains unchanged when we vary the initial mass densities. Of course, the time evolution of $A_2(\tau)$ is not enough and we still need to explore the other important properties of the stellar bar. 

For a better comparison between our models, the 2D projection of the particles in each model is presented in Figures \ref{pos-LPH}, \ref{pos_NLG}, \ref{pos_MOG}, \ref{pos_LIH} and \ref{pos_LHH}. According to the face-on projection of these figures, the bar in the DM models seems to be more elongated than in modified gravity models. For the LHH model, the evolution is completed in a longer period of time. Other features apparent in these figures will be discussed in the next sections.

\begin{figure} 
{\includegraphics[width=0.45\textwidth]{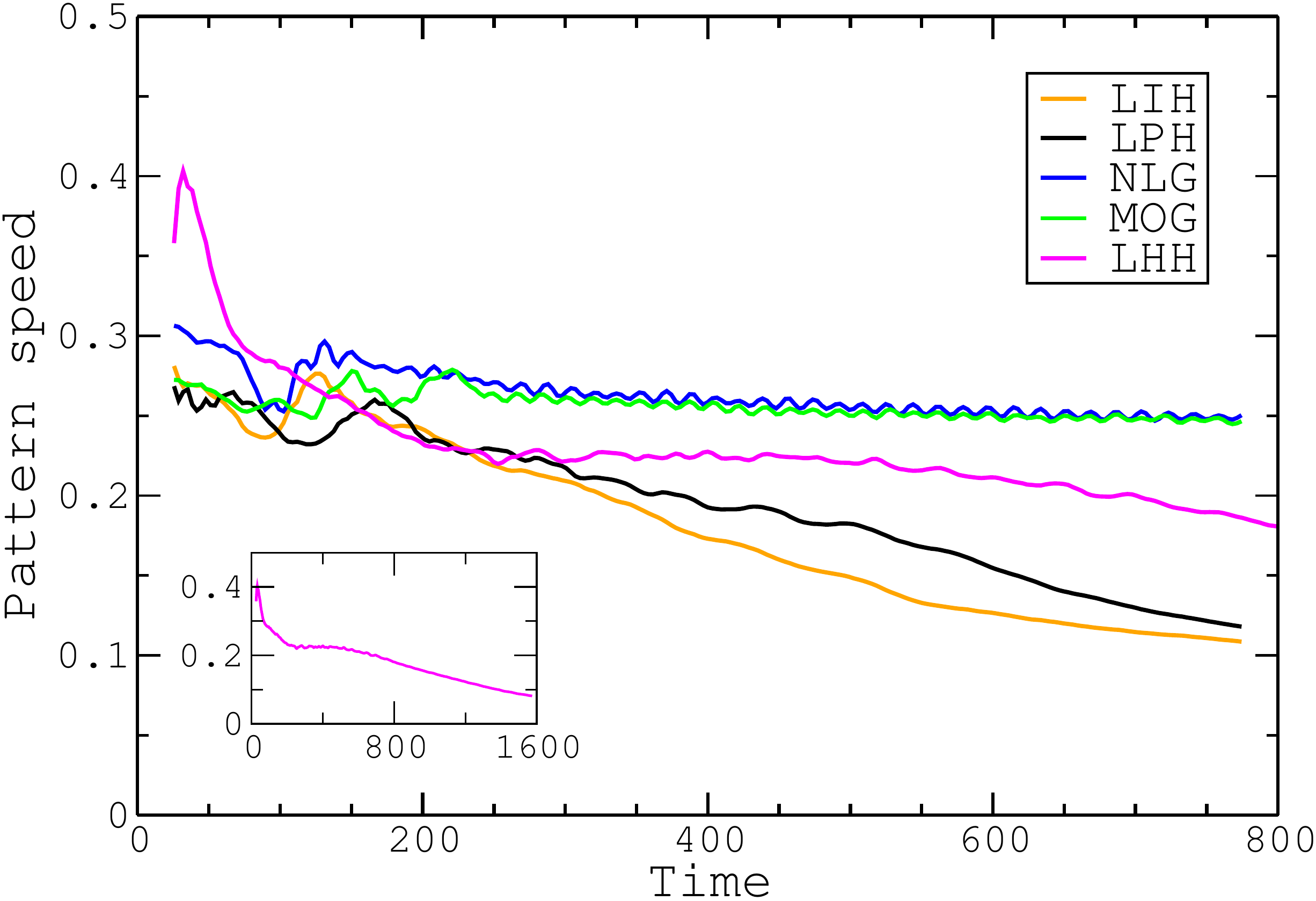}}
\caption{Time evolution of pattern speed for the disk under DM halos LPH (black), LIH (orange) and LHH (magneta), and modified gravity theories NLG (blue) and MOG (green). The cored halo models of LIH and LPH have a sharp drop after $\tau \simeq 200$, while the NLG and MOG curves have small amplitude oscillations about $\Omega_p\simeq 0.25$. The cuspy halo model of LHH starts its final decreasing phase after $\tau \simeq 500$. The subplot shows that it continues its decreasing phase until the end of the simulation. (colors online)}
\label{pat}
\end{figure}

\subsection{Bar Pattern Speed}\label{s.pat}
Another quantity that behaves differently in modified gravity and DM models, is the bar pattern speed. 
Unlike in observations, where the required information should be derived from only one snapshot, in numerical simulations, the evolution of the disk and the rotation of its bar are measured much easier. Therefore, away from the complications of the observational methods, one can simply observe the bar rotation in the simulations to calculate the bar pattern speed $\Omega_p$.
The evolution of this quantity with time is demonstrated in Fig. \ref{pat}. From this figure, it is apparent that the pattern speed in the models of modified gravity is higher than DM models. This quantity decreases with a very slow rate in modified gravity, while it has a significant drop in the DM models. The decline in bar's pattern speed of the cored DM models begins at $\tau\simeq 200$ when the bar strength starts to increase and the second phase in the evolution of the bar amplitude has begun. This is because the halo is responsive and the bar slows down because of the dynamical friction caused by dark matter particles. On the other hand, for the LHH case, similar to the bar amplitude evolution, a constancy phase is visible in the pattern speed, which lasts until about $\tau \simeq 500$ and then the second phase of the drop in pattern speed begins. In this period the dynamical friction starts to act more effectively. However, such behavior is not observed in any mass models of the current and previous simulations of modified gravity \citep{2007A&A...464..517T,2019ApJ...872....6R}. In other words, as expected there is no dynamical friction in modified gravity models. Of course, the dynamical friction experienced by the bar via disk particles is negligible.  Notice that the previously mentioned oscillations also appear in the pattern speed of the modified gravity models. Before moving on to discuss the $\mathcal{R}$ parameter which is of observational importance, let us discuss the LHH model with more emphasis on the amount of the dynamical friction in this model.

\begin{figure} 
\centerline{\includegraphics[width=0.40\textwidth]{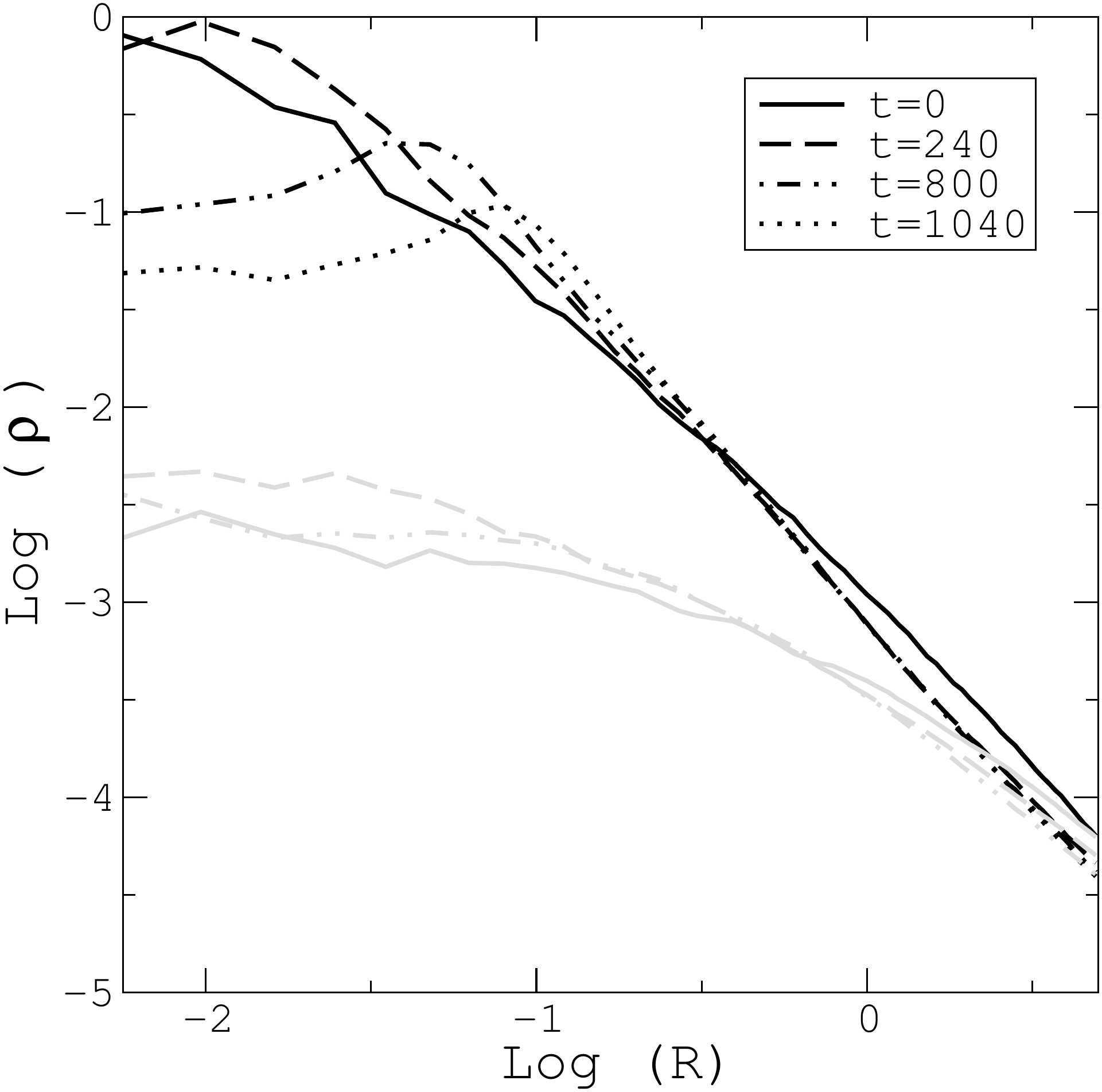}}
\caption{Halo density in the inner parts for different evolution time in the LHH model (black) and LPH model (grey) curves. The initial cusp of the LHH model is transformed to a core according to the secular evolution.}
\label{log-dens}
\end{figure}

\subsubsection{Cusp/Core effect in the LHH model}
To have a better understanding of the behavior of the LHH model, we studied the Hernquist DM halo density during the evolution of the system. As one can see in Fig. \ref{log-dens}, at the beginning of the simulation, there exists a cusp in the inner regions of the LHH model (black curves). As the simulation continues, the secular evolution in the system transforms this cusp to a core that flattens further during the simulation. It is clear from the grey curves in Fig. \ref{log-dens} that such transformation is absent in the initially cored Plummer halo model, which is almost unchanged during the simulation time.

As we already discussed in Sec. \ref{s.bar}, the initial rise in the bar amplitude of the LHH model is followed by a period in which the bar is weakened. According to the projected face-on snapshots of Fig.\ref{pos_LHH}, it is visible that during this period, there is an almost spherical configuration at the center, i.e. a bulge, which might have formed due to the gravitational effect of the cuspy halo. On the other hand, it is well established that a massive bulge has stabilizing effects against global instabilities, for example, see \cite{2001ApJ...546..176S}. This symmetric configuration results in the suppression of the bar amplitude and also reduce the dynamical friction between the halo and disk particles since there is not substantial bulk motion to cause significant dynamical friction. We expect that a moving object inside a medium feels more friction compared to an object which rotates without any linear motion. Therefore, the pattern speed remains almost constant. 
On the other hand, the weak bar activity eventually destroys the cusp, see Fig. \ref{log-dens}. This mechanism has been reported in the literature to address the so-called core-cusp problem \cite{2004ApJ...607L..75E}.  Consequently, the bar starts to grow, as one can see in Fig. \ref{bar}. This directly means that the dynamical friction would rise and the pattern speed will decrease accordingly, see Fig. \ref{pat}.

It is worth mentioning that we also checked another cuspy model of NFW alongside its MOG and NLG counerpart models, however with a different initial rotation curve. From these simulations, we also found similar results for the evolution of the disk under the effect of the cuspy DM halo of NFW. Because of this similarity, we avoid including these simulations for the sake of brevity.

\begin{figure*} 
\centerline{\includegraphics[width=9cm]{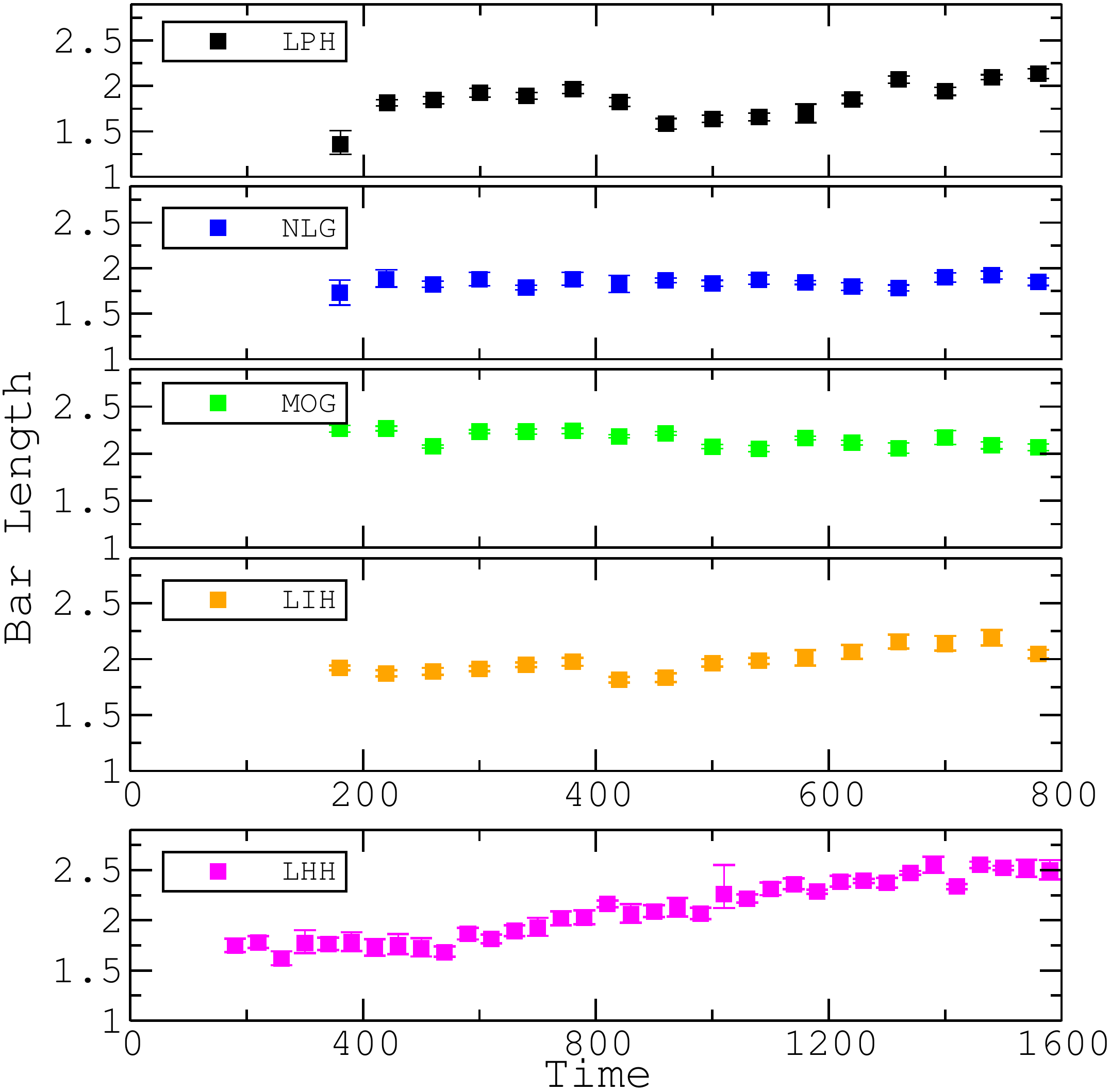}\includegraphics[width=9cm]{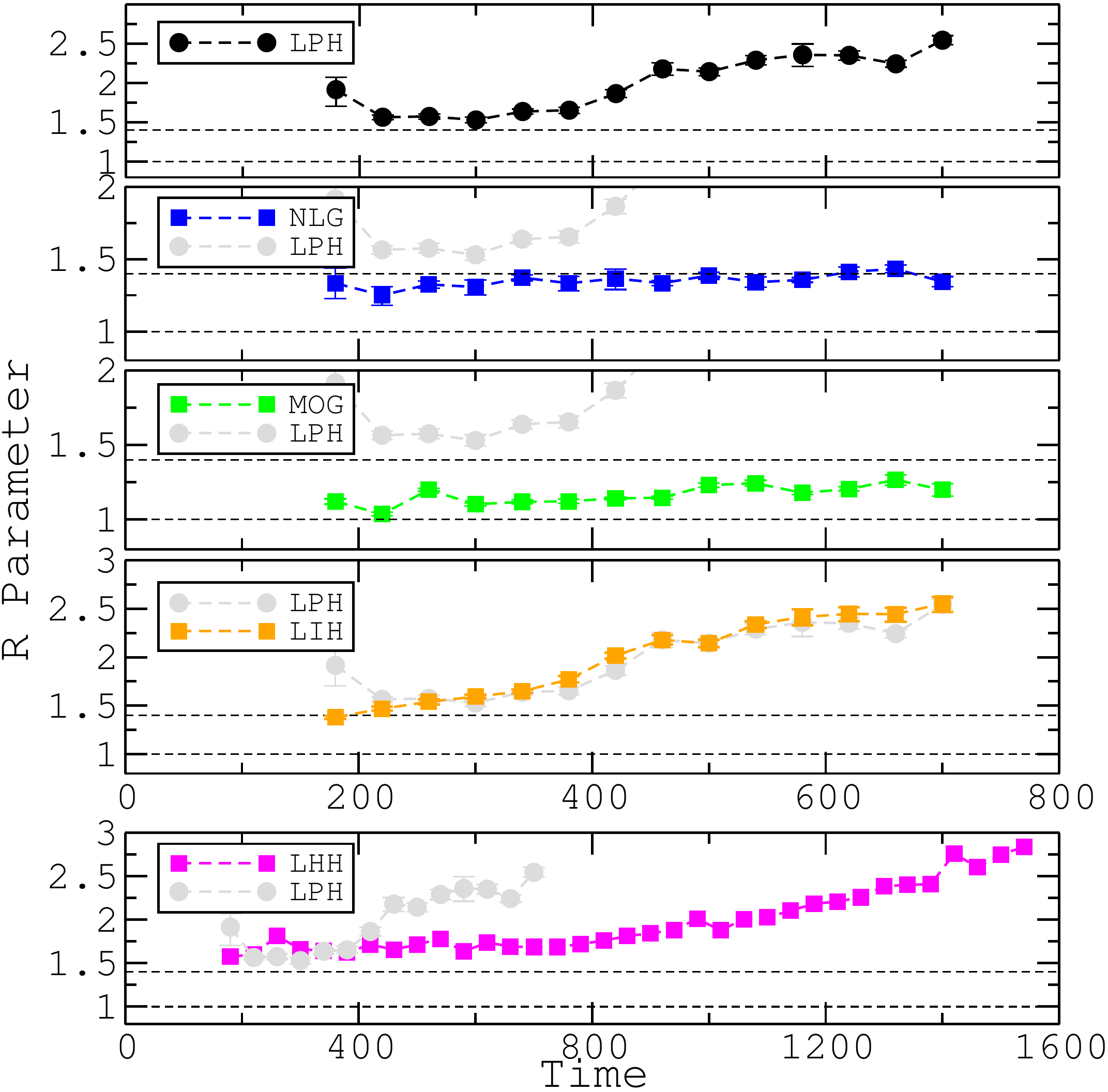}}
\caption{Time evolution of the bar length in the left panel and the $\mathcal{R}$ parameter in ther right panel for our models of modified gravity and DM. Contrary to the DM models that predict slow bars, fast bars are resulted from modified gravity models. The area between the horizontal dashed lines specifies the region of the fast-bar regime. }
\label{rparameter}
\end{figure*}

\subsection{$\mathcal{R}$ Parameter}\label{s.R}
Another useful quantity that is related to the bar pattern speed and has observational importance is the $\mathcal{R}$ parameter, defined as
\begin{equation}
\mathcal{R}=\frac{D_L}{a_B},
\end{equation}
where $D_L$ is the corotation radius and $a_B$ is the bar semimajor axis. Observations indicate that regardless of the Hubble type, bars in disk galaxies appear to be fast rotators with $1<\mathcal{R}<1.4$ \citep{2003MNRAS.338..465A,2015A&A...576A.102A,2019A&A...632A..51C,2020MNRAS.491.3655G}. However, hydrodynamical simulations in the $\Lambda$CDM paradigm report slow bars in disk galaxies with $\mathcal{R}>1.4$ \citep{2017MNRAS.469.1054A}. This is the case also in isolated disk simulations, for example, see \cite{1998ApJ...493L...5D}. There are some proposals to address this discrepancy in the context of the dark matter hypothesis. For example, it is claimed that ultralight axionic dark matter particles suppress the dynamical friction in the galactic scales and consequently leads to fast bars \cite{2017PhRvD..95d3541H}. However, the current debate on this problem in dark matter models has not ended up with a final conclusion. On the other hand, this issue is simply resolved in modified gravity because of the absence of dynamical friction.

To calculate the $\mathcal{R}$ parameter, at the first step one should find the corotation radius, the radius at which the rotational velocity of the pattern is equal to that of the particles (stars) in the disk. Using the pattern speed $\Omega_p$ introduced in the previous section, it is easy to find the corotation radius.

However, calculating bar length $a_B$ is not a straightforward task, neither in real galaxies nor in simulations. Various methods have been proposed to measure this quantity; For example, the visual estimation \citep{1979ApJ...227..714K, 1995AJ....109.2428M}, ellipse fitting to galaxy isophotes \citep{1995A&AS..111..115W}, Fourier decomposition on the surface brightness of the galaxy \citep{1985ApJ...288..438E, 1990ApJ...357...71O, 2000A&A...361..841A} are among these methods that have been widely used in the literature.

In order to distinguish between the behavior of our models regarding $\mathcal{R}$, we used the Fourier decomposition method in finding $a_B$. In this scheme, Fourier components of the intensity ($I_m$) are calculated and the bar ($I_b$) and inter-bar ($I_{ib}$) regions are introduced as 
\begin{equation}
I_b=I_0+I_2+I_4+I_6
\end{equation}
and
\begin{equation}
I_{ib}=I_0-I_2+I_4-I_6,
\end{equation}
respectively, where $I_m$ is the $m$th component of the Fourier decomposition. Finding the ratio of $\frac{I_b}{I_{ib}}$ throughout the disk, \cite{2000A&A...361..841A} show that the outer radius of
\begin{equation}
\frac{I_b}{I_{ib}}> \frac{1}{2}\left[ \left(\frac{I_b}{I_{ib}}\right)_{max} - \left(\frac{I_b}{I_{ib}}\right)_{min}\right] + \left(\frac{I_b}{I_{ib}}\right)_{min}
\end{equation}
results in a good estimation of the bar length $a_B$. According to \cite{2002MNRAS.330...35A}, the error in finding $a_B$ using this method in numerical simulations is less than $4\%$, but reaches to $\simeq 8\%$ for very thin bars. 

Applying the above technique, the $\mathcal{R}$ parameter could be determined for our models.
However, from the projected face-on view of the models, it is clear that there are some stages, especially at the beginning of our simulations, where there exist strong spiral waves in the system. Similar to the bar amplitude, these spiral arms affect the calculations of the bar length and yield a higher value for this parameter. This could result in artifactualy smaller values for the $\mathcal{R}$ parameter. To avoid this effect, we consider time intervals of $\Delta \tau=40$, choose the minimum value of the bar length in each interval and then report the related $\mathcal{R}$ parameter \citep{2020arXiv200305457H}. The results, namely the bar length and the $\mathcal{R}$ parameter, are presented in Fig. \ref{rparameter}. It is clear from the left panels that at the second half of the simulation time in all the dark matter models, the bar length increases with time. On the other hand, from the right panels, we see that the $\mathcal{R}$ parameter also is an increasing quantity. This directly means that the rate of change in the corotation radius of these models is higher compared to the bar length.

As you may see in Fig. \ref{rparameter}, the bar is in the slow regime ( $\mathcal{R}>1$) for all the DM models and also the $\mathcal{R}$ parameter shows a growing trend. According to the previous studies, this behavior is expected because of the dynamical friction between the bar and the dark matter particles. However, in the modified gravity models, namely NLG and MOG, it is obvious that $\mathcal{R}$ remains almost constant in the fast-bar regime ($1<\mathcal{R}<1.4$) and the bar retains its rotation speed. It is apparent that the NLG model results in slightly higher values of $\mathcal{R}$ parameter than in MOG. It should also be mentioned that this property seems to be independent of the adopted disk/halo mass models. As you may see in Fig. \ref{rparameter}, a similar trend in LPH, LIH and LHH models is apparent. Although it should be noted that the delay in the evolution of the LHH model is also seen in the plots of bar length and $\mathcal{R}$ parameter.

Therefore, regarding the results of pattern speed and $\mathcal{R}$, we could conclude that models under the effect of modified gravity show better compatibility to the observational results, in comparison to DM halo models.

\begin{figure*}[!ht]
   \begin{minipage}{\textwidth}
     \centering
     \includegraphics[width=.45\textwidth]{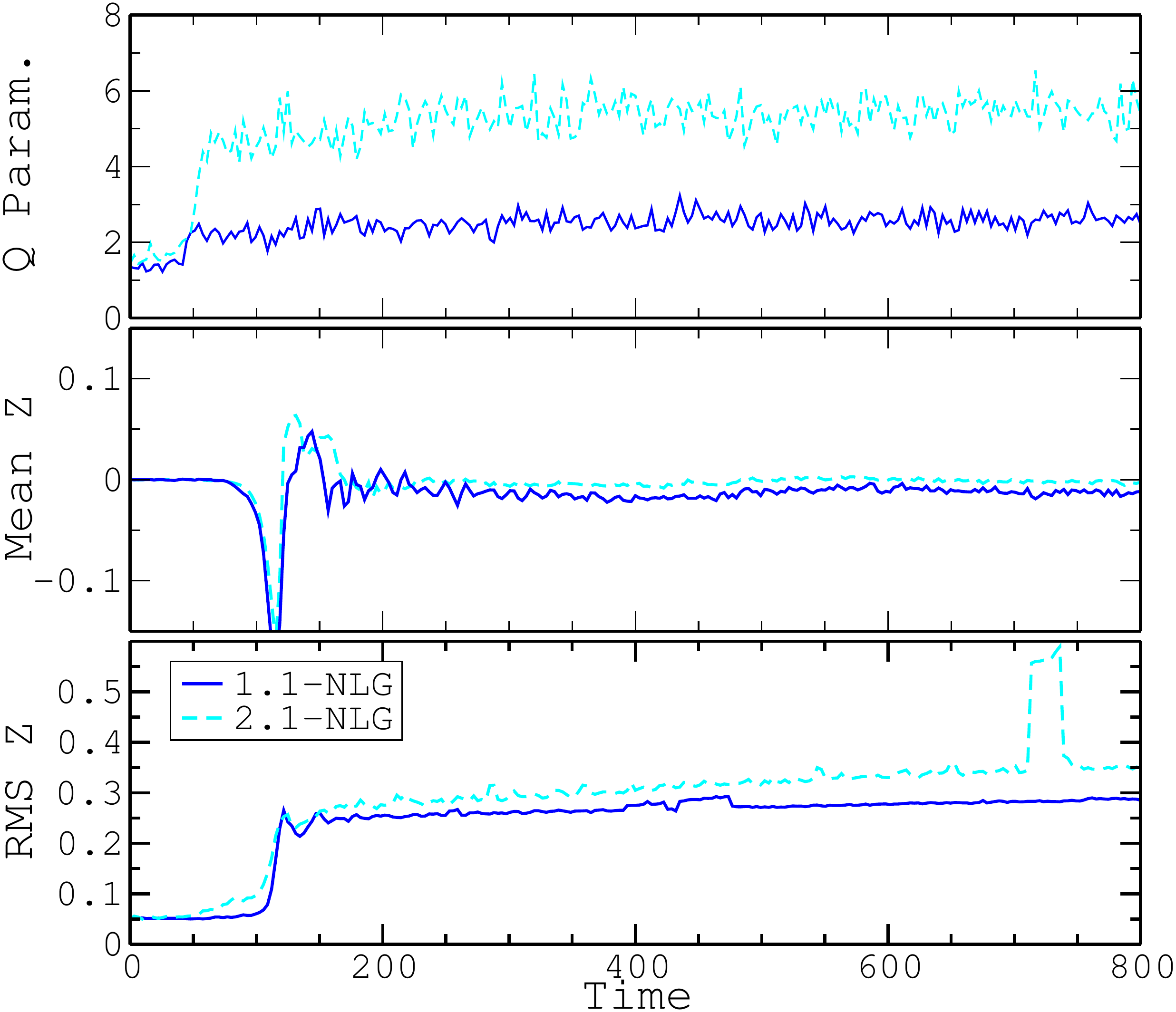}\quad
     \includegraphics[width=.44\textwidth]{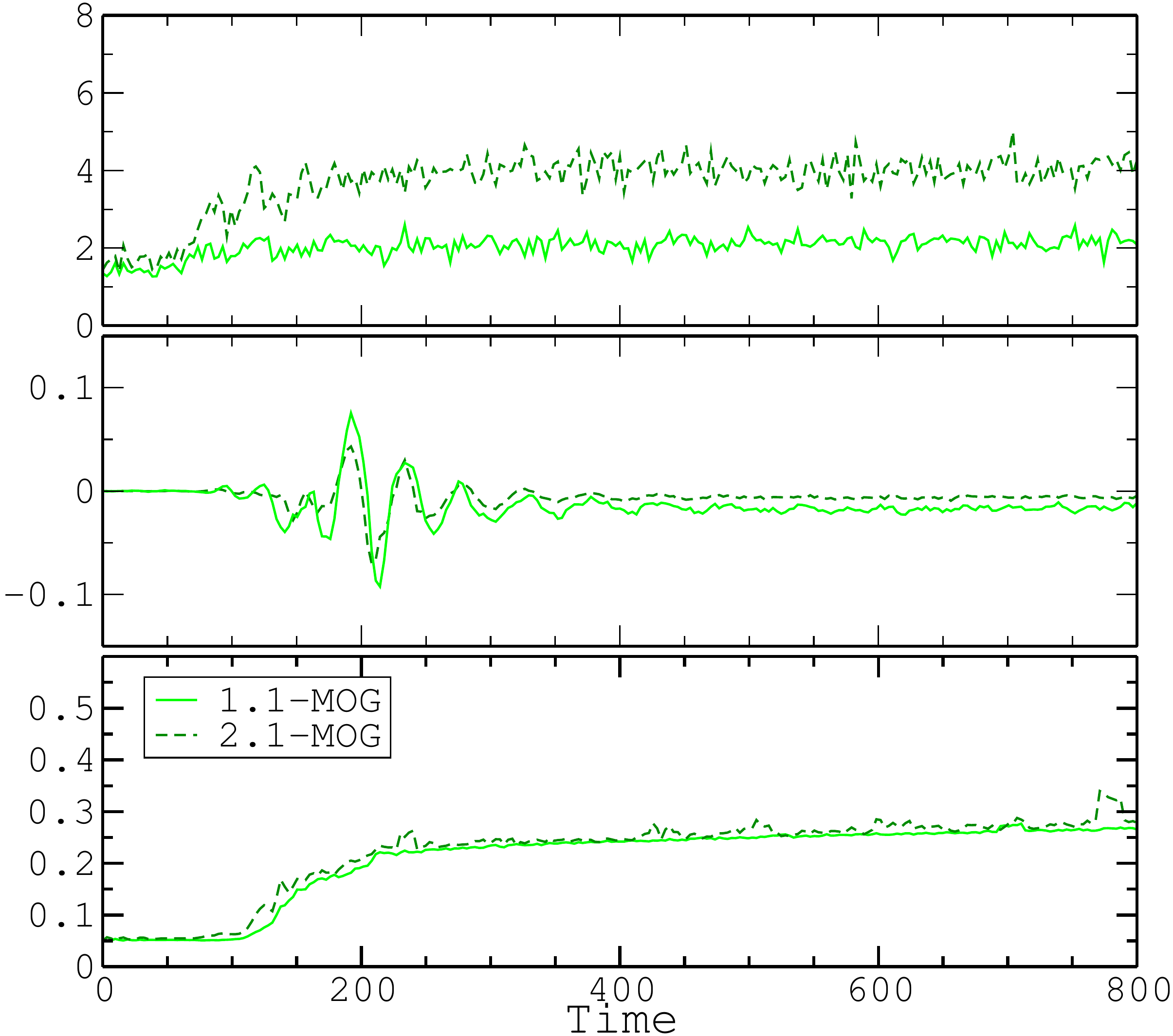}\\
     \includegraphics[width=.45\textwidth]{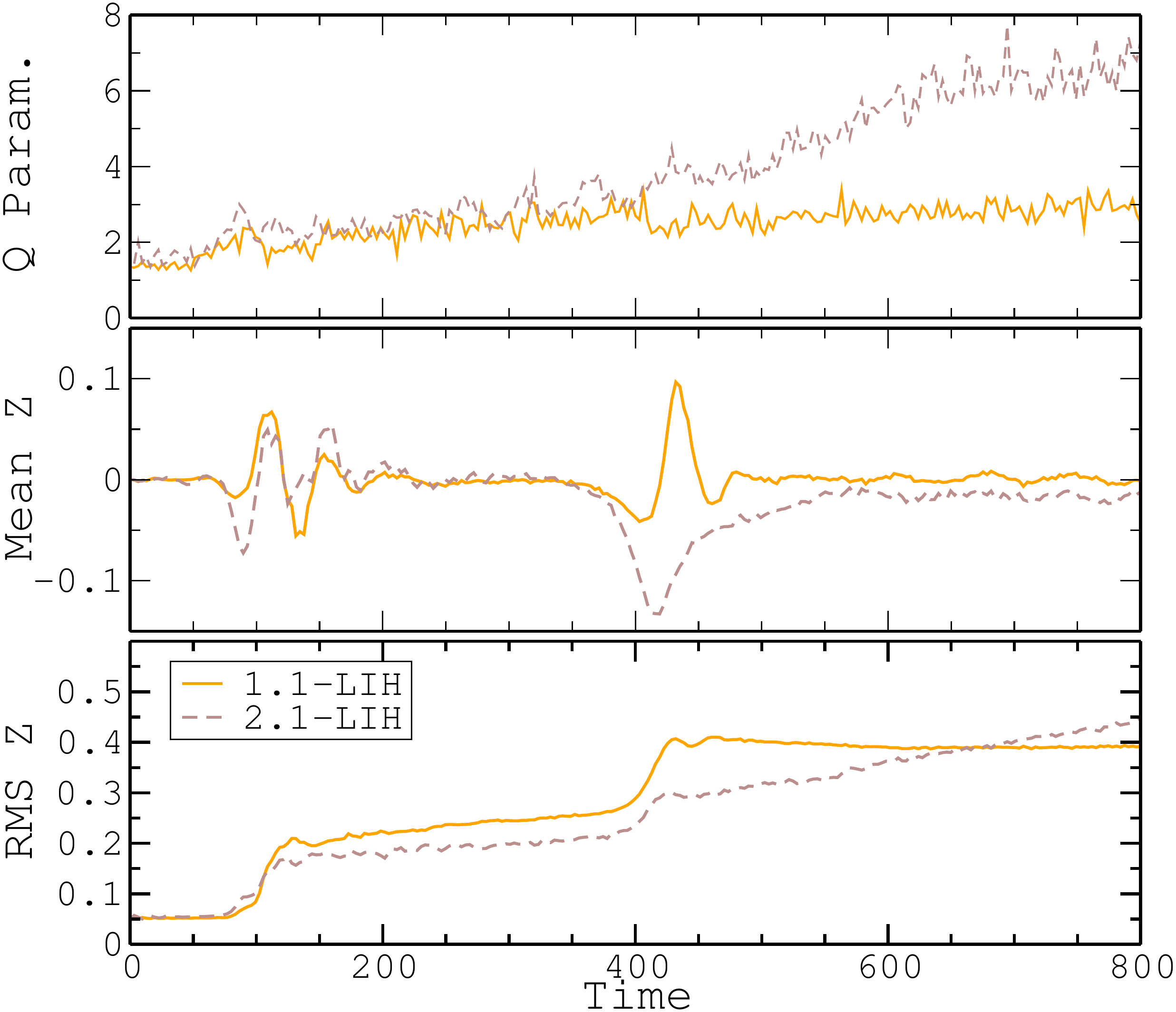}\quad
     \includegraphics[width=.44\textwidth]{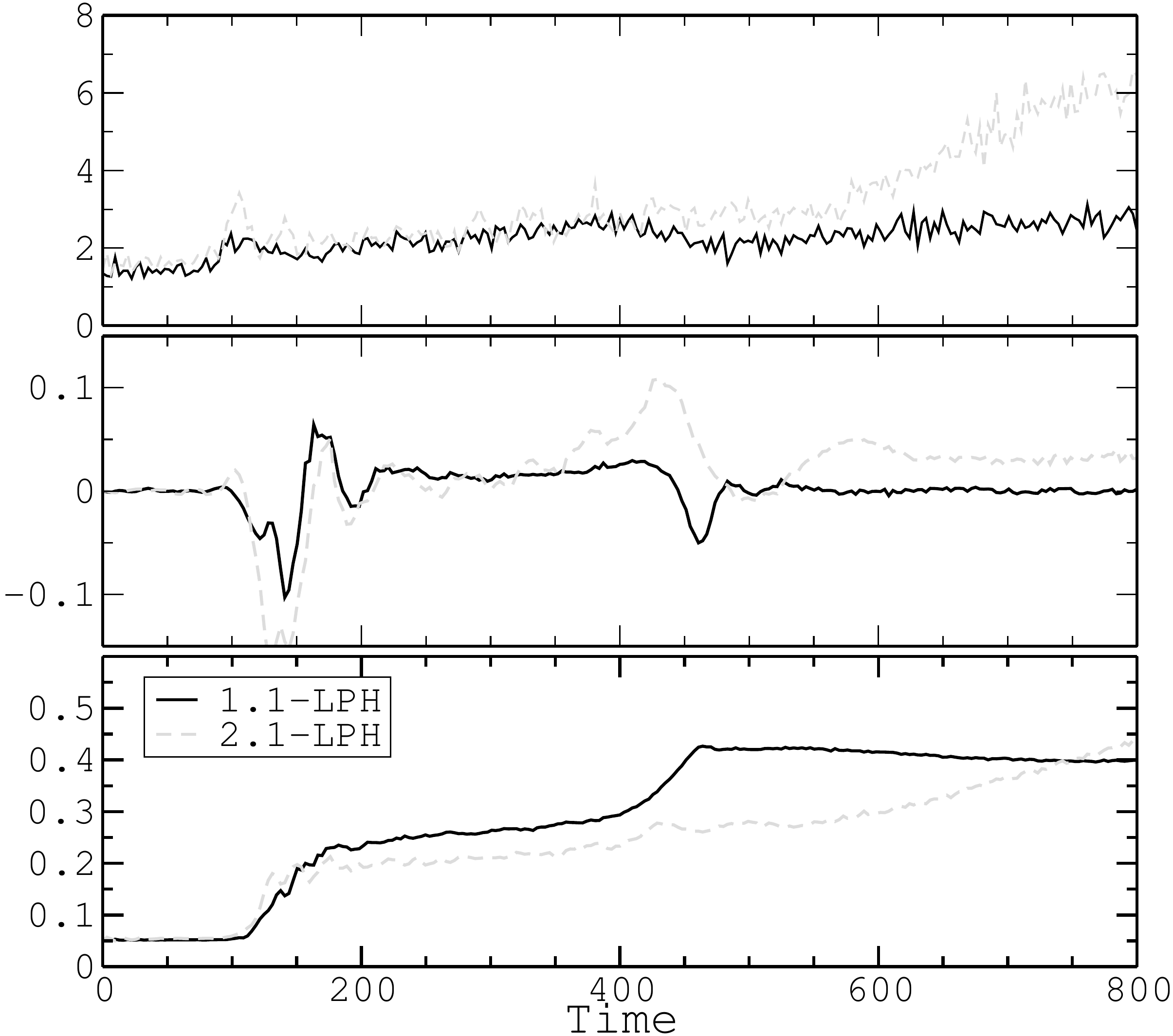}\\
     \includegraphics[width=.45\textwidth]{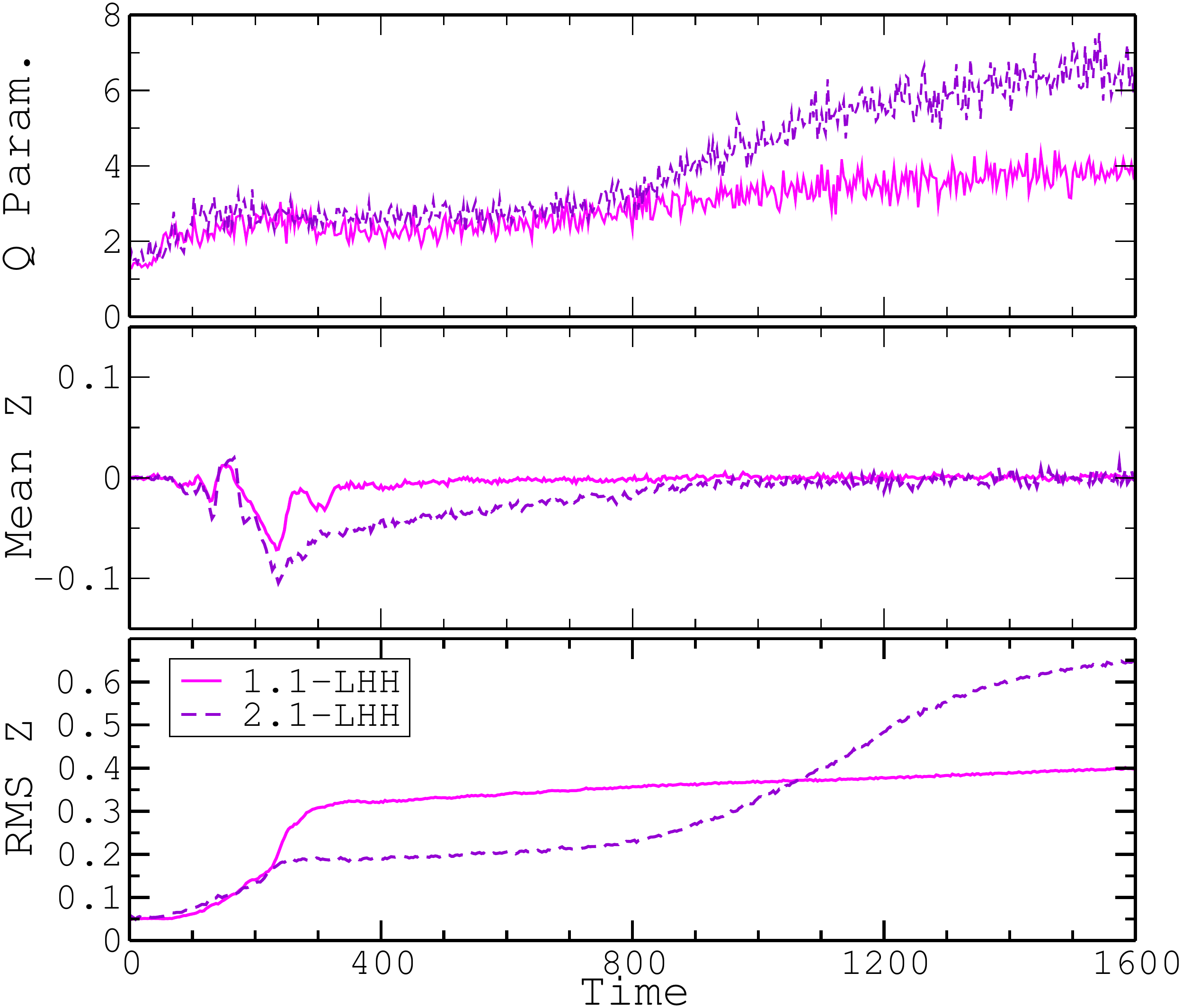}
     \caption{Time evolution of Toomre parameter (upper pannel), vertical mean thickness (middle pannel) and vertical RMS thickness (lower pannel) of the disk for two different radii. Results from NLG (top-left/blue), MOG (top-right/green), LIH (middle-left/orange), LPH (middle-right/black) and LHH (bottom/ magneta) simulations are specified in each part (colors online).  Each quantity is plotted for a smaller radius of $R=1.1$ (solid line) and a larger radius of $R=2.1$ (dashed line). Notice that the LHH model is plotted for a longer period.}
\label{3in1}
   \end{minipage}\\[1em]
\end{figure*} 

\begin{figure*} 
\centerline{\includegraphics[width=0.65\textwidth]{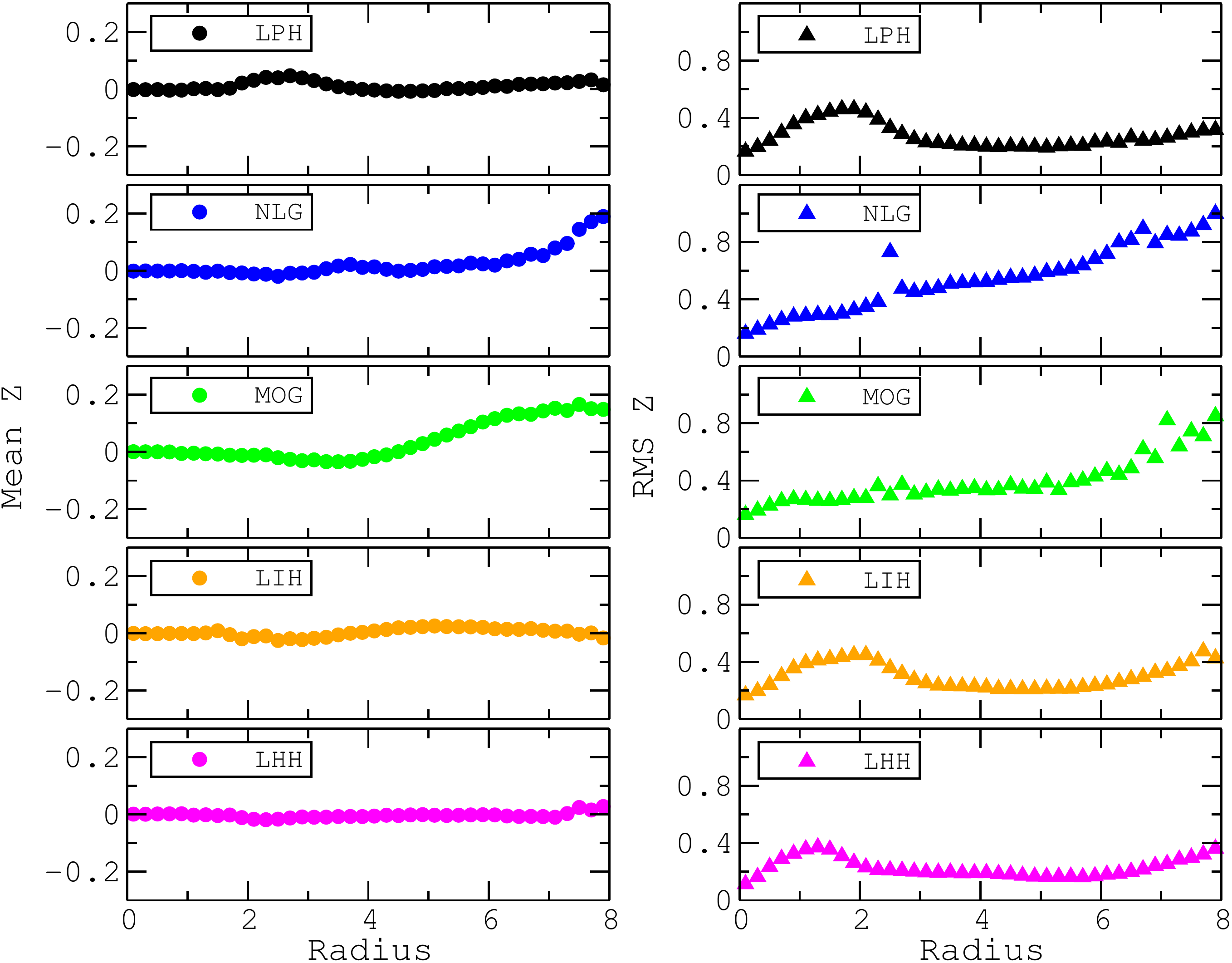}}
\caption{Mean and RMS thickness in terms of radius for each model measured at $\tau=800$.}
\label{z}
\end{figure*}

\subsection{Vertical structure of the Disk}\label{s.vert}
To have a more precise view of the evolution of our models, we studied the disk's behavior in the vertical direction. First, we focus on the inner radii, at which the bar dominates and check the vertical behavior of the models with time. Then we will study the radial behavior of each model's thickness at a final time of $\tau \simeq 800$. The two lower panels of each part in Fig. \ref{3in1}, illustrate the time evolution of the mean $\left<z\right>$ and RMS $\sqrt{\left<z^2\right>}$ thickness of the disk at two inner radii of $r = 1.1$ and $r =2.1$, respectively. In this Figure, the top-left, top-right, middle-left, middle-right and bottom parts present the results for NLG, MOG, LIH, LPH, and LHH, respectively. In all the models studied in this work, except the LHH model, growth in RMS thickness is visible at $\tau \simeq 100$. Similar to the results of the previous sections, the RMS thickness in the LHH model shows delay in its evolution and starts its initial increase at about $\tau \simeq 200$. However, in comparison to MOG and DM models, NLG shows a sharper and more distinct increase, while for the other models, the growth happens at a slower rate. This is compatible with the higher initial bar growth rate of NLG which was discussed in Sec. \ref{s.bar}. It should be noted that the evolution of the vertical behavior of the disk is linked to the bar strength evolution. Comparing the results of the RMS thickness to bar amplitude in our modified gravity models, it is clear that exactly after the buckling instability, i.e., $\tau\simeq 150$ in NLG and $\tau\simeq 200$ in MOG, the bars enter the oscillatory regime around a constant value. On the other hand, a second step-like behavior is visible in both LIH and LPH models at $\tau \simeq 400$. Such a second step-like behavior is not present in the RMS curve of $r= 1.1$ for the LHH model, however, a smooth growth in $r=2.1$ is visible at around $\tau \simeq 800$, which seems to be compatible to the slight change in the slope of the bar amplitude curve after this time.

It should be stressed that each step-like behavior in RMS thickness can be interpreted as the manifestation of the buckling instability. During this instability, the bar is weakened as the result of vertical resonances \citep{1991Natur.352..411R}. It is interesting that in our dark matter models of LPH and LIH this instability happens twice. The first buckling happens at the same time when there is a tangible reduction in the bar amplitude.
However, it should be noted that for all the DM models, the first buckling is stronger, and its effect on the bar amplitude is more explicit. For example, the RMS thickness of the LIH model at $r=1.1$ varies from $0.5$ to about $2$ during the first buckling, i.e., enhancement by a factor of $4$, but grows only from about $2.5$ to $4$, i.e. by a factor of $1.6$, in the second buckling. Consequently the second buckling is not accompanied by a rapid reduction in the bar magnitude. Regarding the fact that $A_2$ is measured using the projected position of the particles in the $x-y$ plane, in the second buckling the bar thickens in the vertical direction and there is not a substantial change in its projected width and length measured in the $x-y$ plane. 

 From Fig. \ref{3in1}, it is also apparent that in DM models, the trend of RMS thickness at larger radii (dashed curves) in the second half of the simulation differs from the smaller radii (solid curves) and grows with an almost constant rate, while it stays almost constant at smaller radii. This could be because a peanut shape is formed in our DM models. Although the disk thickens during the first buckling, the peanut shape appears more explicitly after the second buckling in LPH and LIH models. However in the LHH model, although there is a smooth second buckling, there is a vivid peanut showing up after the first buckling. To see this fact more clearly, refer to the plots of the particles' projected position in different planes of each model in figures \ref{pos-LPH}, \ref{pos_NLG}, \ref{pos_MOG}, \ref{pos_LIH} . In modified gravity models there is only one buckling and although we see a sudden increase in the RMS thickness, there is no explicit two-fold symmetric peanut.

The mean thickness, which is presented in the middle panel of each part in Fig. \ref{3in1} is a measure for the asymmetric distribution of the particles around the $z=0$ plane. A sudden change around the $z=0$ plane in the $\left< z \right>$ diagram happens at the same time as the growth in RMS thickness. 

The other quantity related to the disk's stability is the Toomre parameter $Q$, for which the time evolution has been illustrated in the upper panel of each part in Fig. \ref{3in1}. We should recall that, to avoid the disk from local collapse and instability, the condition of $Q>1$ should be fulfilled. From Fig. \ref{3in1}, it could be seen that for all of our models, the $Q$ parameter in small radii has a lower value in comparison to larger radii. Started from $Q\simeq1.5$ at $\tau=0$, this parameter in MOG and NLG increases rapidly to $Q\simeq 2$ in $R=1.1$ and remains almost constant until the end of the simulation. However, in the same radius in LIH and LPH models, $Q$ starts from $1.5$ but reaches to about 3 until the end of the simulation. On the other hand, the $Q$ parameter in the LHH model behaves similar to the other DM models at this radius until $\tau \simeq 800$ and then continues to grow smoothly to about 4 until the end of its evolution. 

\begin{figure*}[!ht]
   \begin{minipage}{\textwidth}
     \centering
     \includegraphics[width=.45\textwidth]{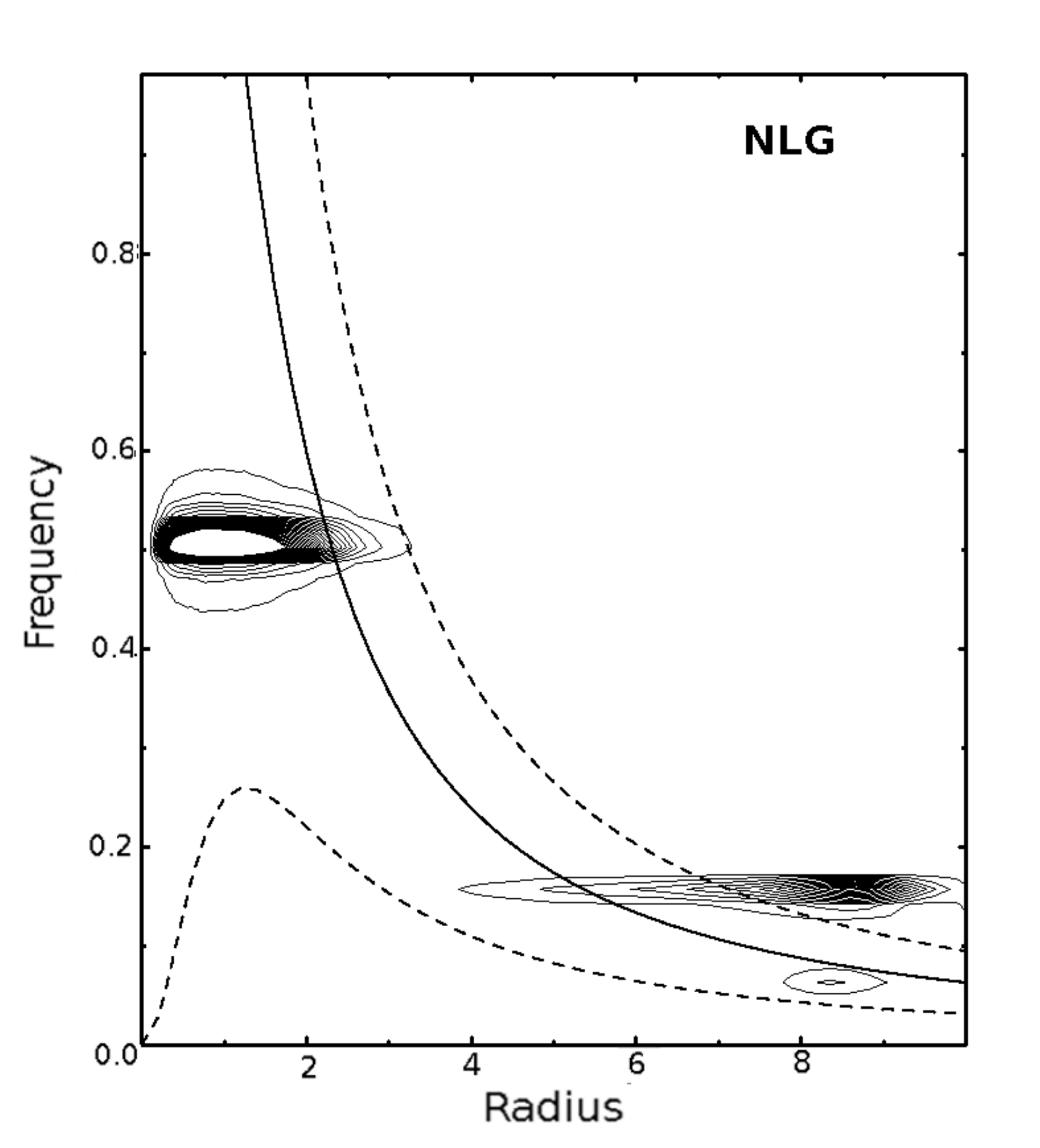}\hspace{0.3cm}
     \includegraphics[width=.45\textwidth]{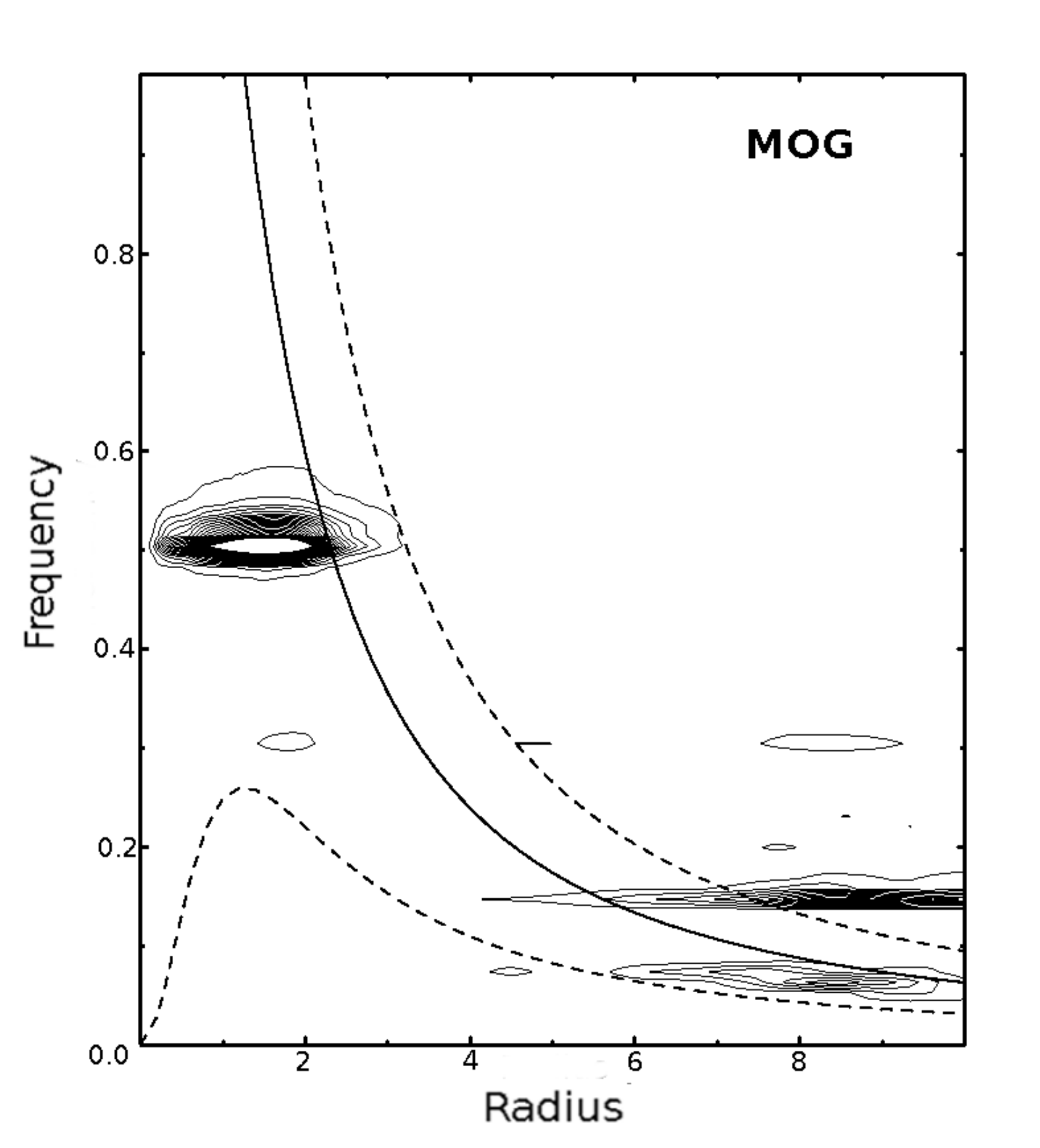}\\\vspace{0.5cm}
     \includegraphics[width=.448\textwidth]{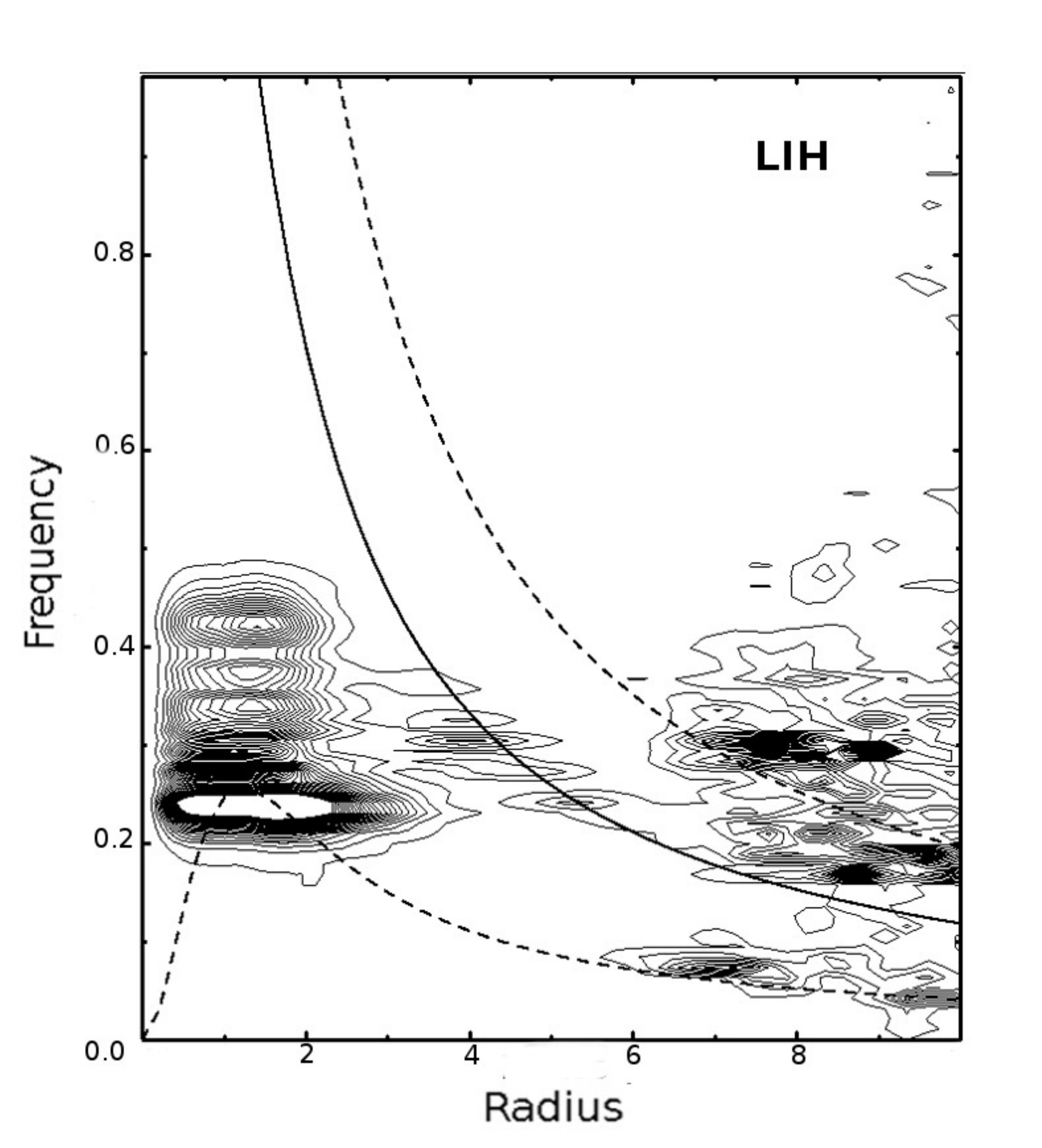}\hspace{0.3cm}
     \includegraphics[width=.45\textwidth]{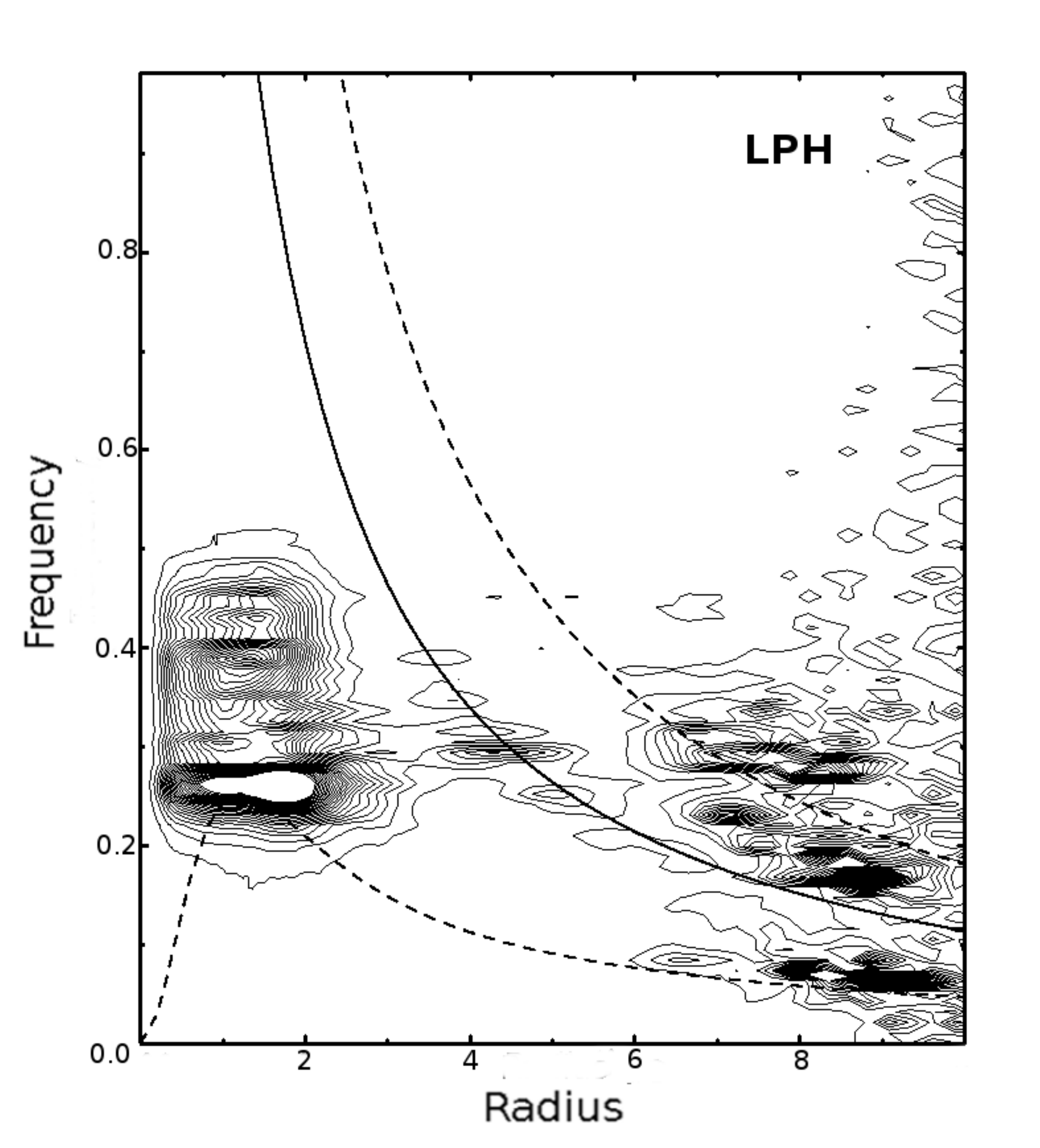}
     \caption{\textit{top-left panel:} The contours of the power spectrum of NLG model for $m=2$. \textit{top-right panel}: power spectrum for MOG model. \textit{bottom-left panel:}  power spectrum for LPH, and \textit{bottom-left panel:} shows the corresponding spectrum for LIH model. For all the models the power spectrum in evaluated for the time interval $200<\tau<800$.}
\label{ps}
   \end{minipage}\\[1em]
\end{figure*} 

\begin{figure*} 
 \centerline{\includegraphics[width=0.45\textwidth]{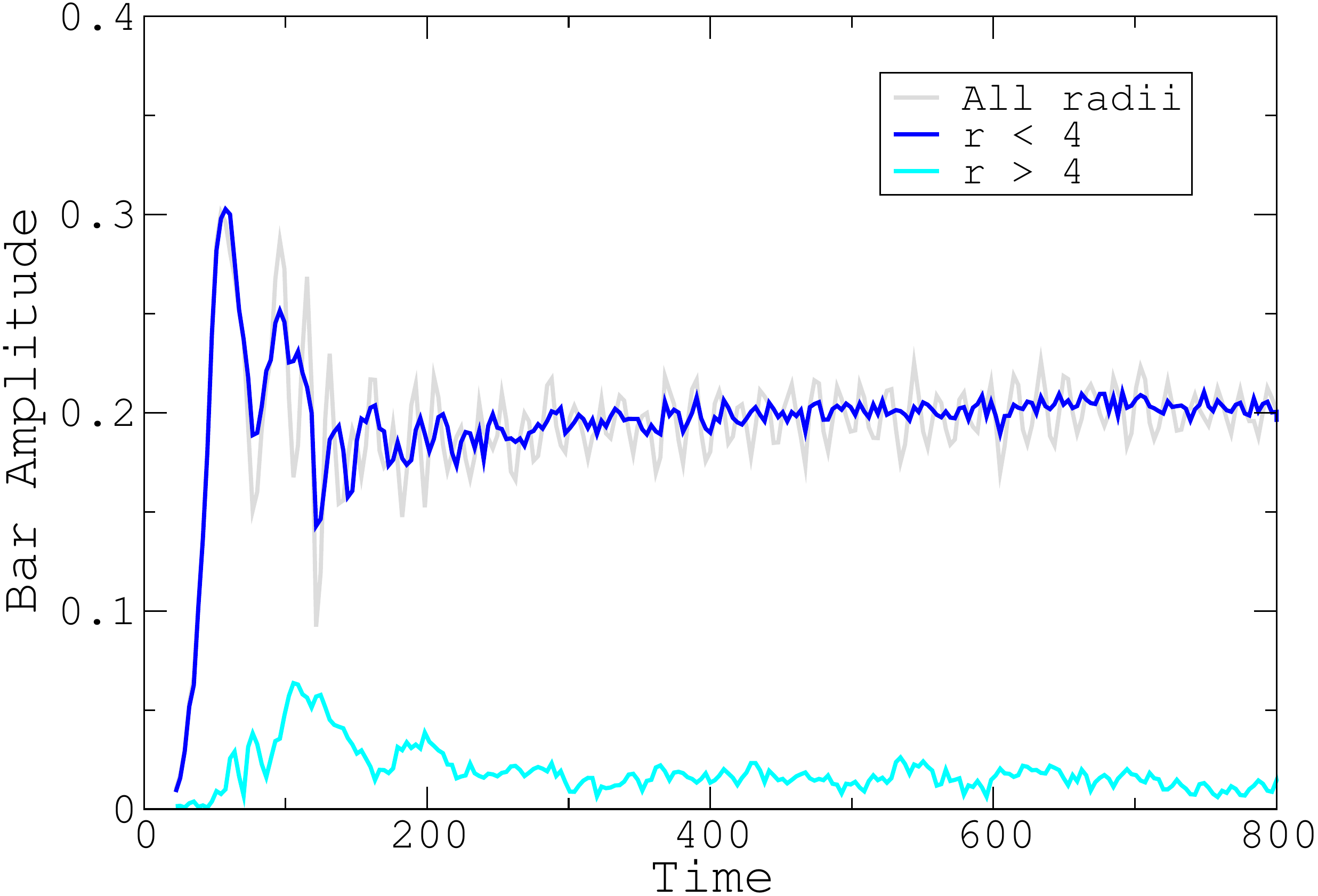}\includegraphics[width=0.45\textwidth]{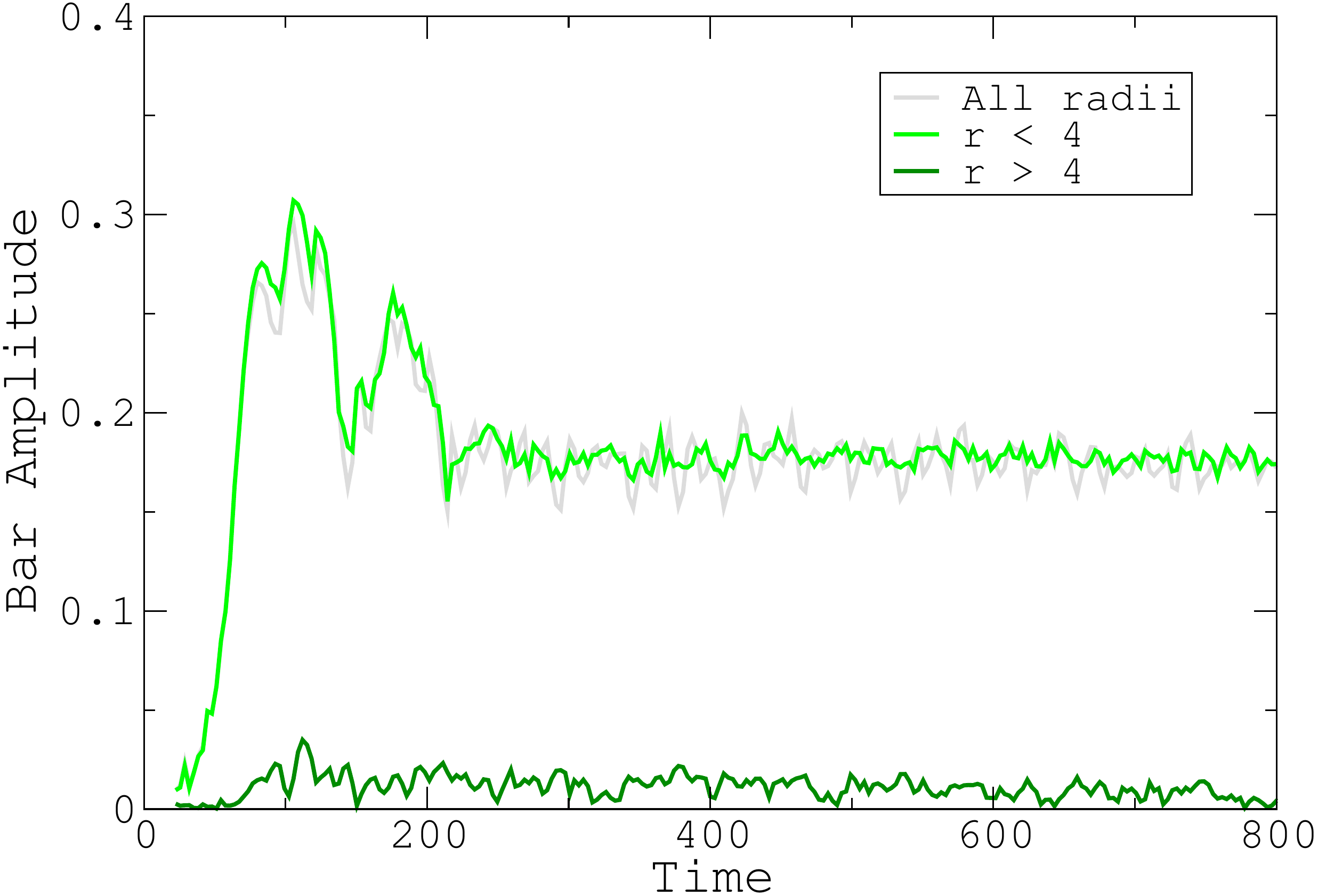}}
\caption{Reduction of the oscillations in bar amplitude by eliminatingt the effect of the modes in the outer radii of the NLG and MOG disks in the left and right panels, respectively.}
\label{figinout}
\end{figure*}

Modified gravity and DM models show another difference at larger radii. Although they start from the same initial value for $Q$, NLG and MOG show a sharper growth at an earlier stage of the evolution. In the NLG model, this growth happens at about $\tau \simeq 50$ and reaches $\simeq 5$ and remains constant until the end of the simulation. Also in the MOG model, we see that the growth starts at $\tau \simeq 50$, however, the growth rate is not as fast as in NLG and lasts until $\tau \simeq 200$, where the $Q$ parameter reaches to an almost constant value $\simeq 4$. On the other hand, in the DM models, this growth is smoother and starts from the second half of the simulation ($\tau \simeq 400$ for LPH and LIH models and $\tau \simeq 800$ for LHH model), where the second buckling instability happens in the disk. The final value of the $Q$ parameter at the end of simulation reaches to $\simeq 7$ in all the DM models.

To have a better view of the overall behavior of the disk in the vertical direction, we plotted the mean and RMS thickness in terms of radius for all the models at $\tau=800$ in the left and right panels of Fig. \ref{z}, respectively. The RMS thickness in LIH and LPH models is higher than modified gravity models only at radii about $ 1\lesssim R \lesssim 2$. This peak in RMS thickness is a result of the peanut shape that occurs in the DM models. The peak in the LHH model at $\tau=800$ has a lower value and happens at a smaller radius.
On the other hand, the RMS thickness in MOG and NLG models rises almost smoothly from the inner to outer radii and in most of the disk has higher values than in DM models. In other words, ignoring the inner parts, the disks are thicker in modified gravity models.

From Fig. (\ref{z}, left), it could also be seen that the mean thickness in NLG and MOG shows deviation from zero. This means that, in accordance with the conclusion of \cite{2007A&A...464..517T}, our MOG and NLG models are flared and warped. This could also be seen in the edge-on projection of the disks (Fig. \ref{pos_NLG} and \ref{pos_MOG}).

It is necessary to mention that the vertical behavior of the modified gravity models considered in this work shows some similarities to MOND \citep{2007A&A...464..517T}. On the other hand, there are some differences. For example, the peanut shape explicitly appears in MOND simulations presented in \cite{2007A&A...464..517T}, while we do not confirm its existence in NLG and MOG. These differences are related to the distinct nature of these models. There is a fundamental acceleration $a_0\simeq 10^{-10}m/s^2$ in MOND beyond which these deviations from Newtonian dynamics appear. On the other hand, NLG's corrections originate in the nonlocal features of gravitation. There is no fundamental acceleration scale in this theory. The field equations are also different in these theories. In other words, although the gravitational potential $\Phi$ satisfies a linear differential equation in NLG and MOG, MOND's field equation is nonlinear. Therefore, in principle, we do not expect the completely same behavior for them in galactic dynamics. Further investigations on the comparison between these modified gravity models are required to reveal their viability using N-body galactic simulations.

All the discussion presented in this subsection can be reduced to the fact that at the inner radii, in which the bar is dominating, modified gravity models lead to disks with smaller thickness, i.e., the bars in DM models seem to be thicker and show stronger peanut configurations. However, the disk in modified gravity models at larger radii is more warped and flared and shows an overall higher thickness in comparison to the DM models. This is of great importance in the sense that it implies that the morphology of the disks could be different in modified gravity compared to dark matter models. It is necessary to mention that in a similar phenomenology, it is recently claimed that the vertical structure of the galaxies could help to discriminate between modified gravity and dark matter \cite{2019PhRvD.100h3009L,2019arXiv191112365L}.

As the final remark in this section let us recall that the above-mentioned fact dealing with the thickness of the disks at small radii, has also been reported in \cite{2018ApJ...854...38R,2019ApJ...872....6R} for different mass profiles. Therefore, regarding the main purpose of this paper, it seems that the emersion of disks with smaller thickness at small radii compared to dark matter models is a model-independent feature in modified gravity models. Let us add our new finding to the previous results: disks at large radii are thicker in modified gravity models. In other words, ignoring the central region, the galactic disks seem thicker in modified gravity models compared to the standard picture.

\subsection{Power Spectrum}\label{spectrum}
In Fig. \ref{ps} we have plotted the contours of the power spectrum with respect to the radius in the second phase of the simulations $\tau>200$. This figure helps to understand the oscillating behavior of the bar magnitude. It should be emphasized that although we use different modified gravity model parameters $\alpha$ and $\mu$ compared to \cite{2018ApJ...854...38R}, we see the same oscillatory behavior reported in the above-mentioned paper. In Fig. \ref{ps}, the top-left, and top-right panels belong to NLG and MOG respectively. In both models, we see at least two distinct frequencies in the power spectrum. This means that there are two density waves with different frequencies propagating on the surface of the disk. Naturally, this yields oscillatory behavior in the physical properties of the bar. Interestingly the frequency of these waves in NLG and MOG is close to each other. This is expected in the sense that the pattern speed of the bar in these models is also close to each other.

In the bottom-left and bottom-right panels, we have shown the power spectrum for LIH and LPH models. The vertically aligned contours directly mean that there is no constant frequency in the system. In other words, the frequency of the bar is varying with time. We know that it is a decreasing function with respect to time because of the dynamical friction. This feature is seen in both DM models. It should be noted that the noisy concentration of the contours in large distances does not necessarily mean that there are several density waves as the main concentration of the baryonic matter is limited to smaller radii, as you may see in Fig. \ref{pos-LPH} and \ref{pos_LIH}. However, as also reported in previous studies, disks in modified gravity expand to larger radii. Therefore, the analysis of the power spectrum would be fully relevant at these radii.

As another test on the oscillatory behavior in our modified gravity models, we calculated the bar amplitude in the inner region of the disk, i.e. $r\lesssim 4$, where the bar mode dominates. As mentioned earlier, it is clear from Fig. \ref{ps} that there are two distinct modes in radii $r < 4 $ and $r > 4$ in these two models. The results are compared to the bar amplitude that was previously calculated for all the disk particles (grey curves) in Fig. \ref{figinout}. The value of $m=2$ mode in the outer disk is also presented in this figure. All the values are scaled to the total number of the particles (mass) in each model. As it is clear in Fig. \ref{figinout}, the amplitude of the oscillations in the NLG and MOG models considerably changes by considering the particles inside $r\simeq 4$. It could be concluded that the oscillations in the bar amplitude and therefore in pattern speed are directly related to the existence of the second mode in the outer disk.  

\begin{figure*} 
 \centerline{\includegraphics[width=0.45\textwidth]{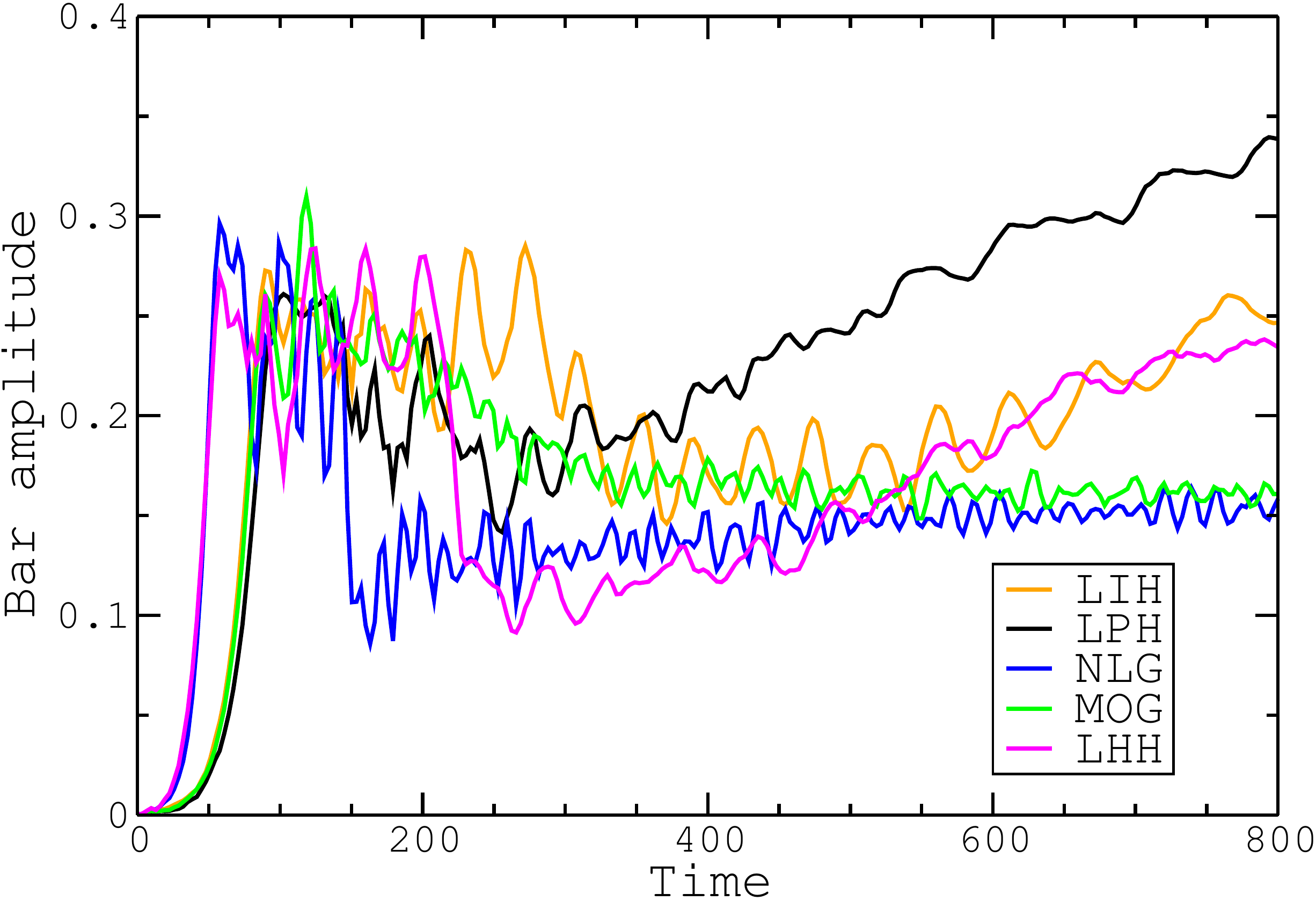}\includegraphics[width=0.45\textwidth]{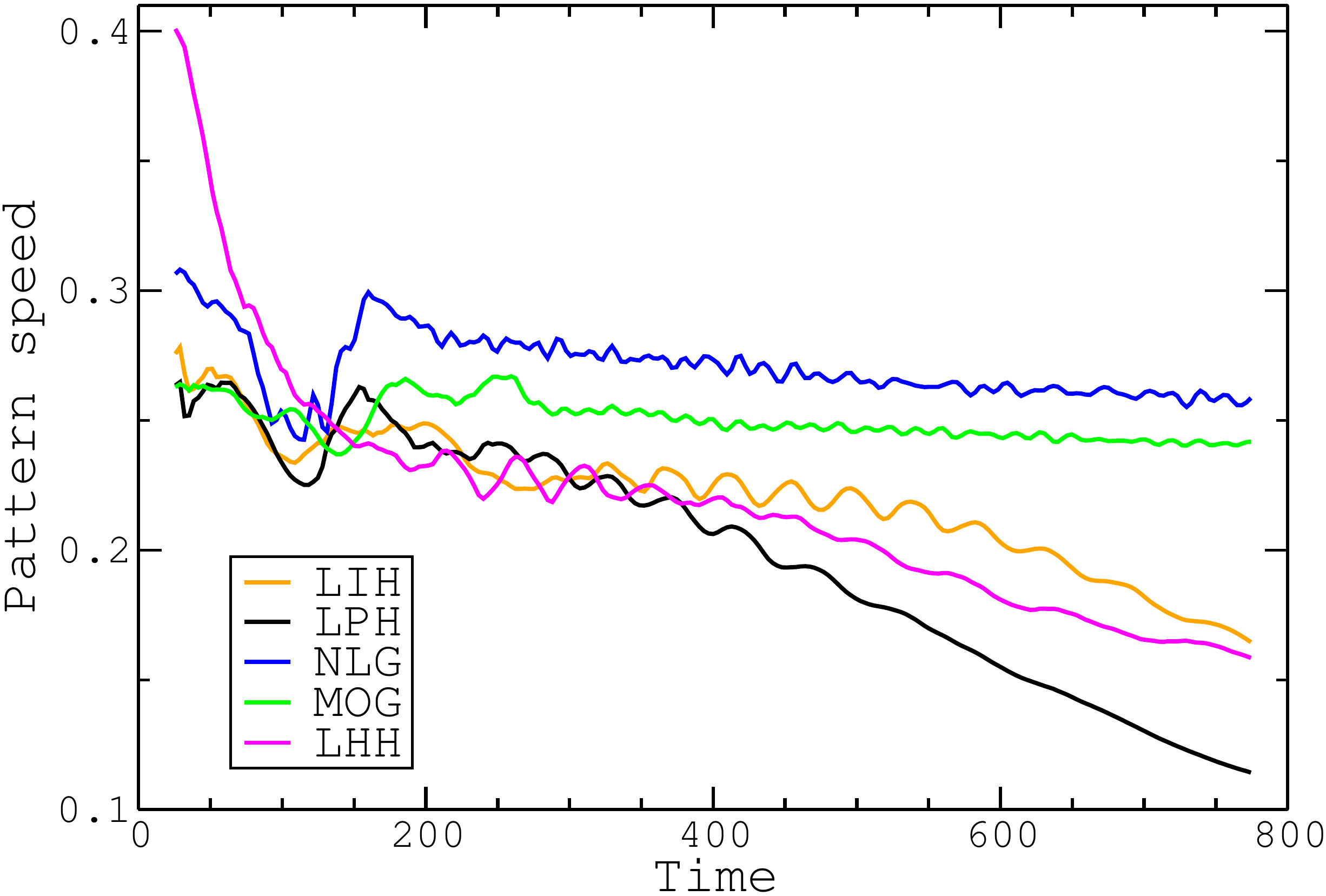}}
\caption{Results of the evolution of bar amplitude (left) and pattern speed (right) for our models with $N=10^7$ particles.}
\label{10million}
\end{figure*}

\subsection{Tests on The Reliability of The Results}\label{tests}
To ensure that the results do not suffer from numeric artifacts, we have performed some tests. The conservation of energy is checked to be less than 2 percent until the end of the simulation. To make sure the main results are not affected by shot noise, we checked the results by changing the number of particles from 2 to 10 million. For illustration, some of the results are reported in Fig. \ref{10million}. As you may see in Fig. (\ref{10million}, left), similar to the models with 2 million particles the initial peak in the bar amplitude happens earlier in NLG models. A similar behavior is seen in the LHH model, however, the peak in LHH is slightly lower than the NLG model, which is compatible with the result of the low-resolution simulation. This initial increase happens later and with a slower rate in the MOG model, and then in the DM models, as was also seen previously. On the other hand, the drop of bar amplitude in the MOG model with $N=10^7$ is smoother, however, at later stages, both modified gravity models reach an equal amount of bar amplitude again. Furthermore, the second phase of the increase in the bar amplitude is clearly seen in all the DM models. It should be noticed that the constancy phase in the evolution of the LHH model is shorter and the increasing phase starts faster in comparison to the low-resolution simulation. Fig. (\ref{10million}, right) illustrates the evolution of pattern speed in our models with $N=10^7$ particles. As it is apparent, the value of pattern speed in our modified gravity models is considerably higher than in DM models. On the other hand, the fast drop in the LIH model, and the convergence of the NLG and MOG models happens later in comparison to the $N=2\times10^6$ case. However, it is clear that the main differences in the characteristics of DM/modified gravity models remain unaltered. It should be mentioned that changing the number of particles, such delay in the appearance of features during the evolution of disks, has also been found in previous studies \cite{2017MNRAS.468.4450G}. 

Altering the time-step from $0.01$ to $0.005$ leads to the same results as in our main simulations. This means that the time evolution is already converged and we use a suitable choice for the time step. Furthermore, considering a higher resolution by changing the mesh dimensions to be $240 \times 256 \times 135$, and changing lscale from $12.5$ to $14.5$, keeps the main results un-altered.

\section{Discussions and conclusions}\label{conc}
In this manuscript, we made a comparison between the evolution of disks under the effect of two modified gravity models, NLG and MOG, and three DM halo models consisting of a live Plummer halo (LPH) and a live isothermal halo (LIH) which are considered as cored DM halos, and a live Hernquist halo (LHH) which has a cusp at the beginning of the simulation. The same baryonic disk, namely the Kuzmin-Toomre profile, was adopted for all five models and the initial conditions of the models were constructed in a way that they start their evolution from a similar state. Therefore, the initial rotational velocity, velocity dispersions and the initial Toomre parameter of the models selected to be as identical as possible. However, the dynamical evolution of the models does not follow the same track. 

The formation of the stellar bar in the NLG model is faster and has a higher amplitude.  However, after the buckling instability, it experiences a rapid reduction and remains almost constant until the end of the simulation. The same trend is followed by the MOG model, however, the rate of bar formation is slower and it has a lower amplitude. On the other hand, it has been illustrated that for the DM models, the initial rise in the bar amplitude is lower and slower than the NLG case but after a drop in its value (buckling instability), it starts to grow again. The evolution in the low-resolution LHH model is slower in comparison to the other halo models. The suppression of the bar amplitude continues for a longer period. However, In about half their simulation time, all the DM models undergo another buckling. The outcome of this buckling instability in the LPH and LIH models is the appearance of a stellar peanut configuration. Then, they start growing again at an almost constant rate.  According to this behavior and regardless of the initial mass profiles, we could conclude that the system under DM influence, in comparison to modified gravity models under consideration, would result in stronger bars after a limited evolution time. It should be mentioned that, although the second buckling has not been observed in our previous simulations with the EXP disk profile, the higher thickness of the disks in DM models at the inner regions, in comparison to modified gravity models is independent of the adopted disk/halo model.

The evolution of the pattern speed in our four models also shows a resembling trend in NLG and MOG, which is considerably different from all the DM models. The pattern speed of the bar in modified gravity models has some initial fluctuations and then shows small oscillations around an almost constant value. However, in DM models, it starts falling with an almost constant rate after $\tau\simeq 200$ for the LIH and LPH models and $\tau \simeq 500$ for the LHH model, which continues to the end of the simulation. We conclude that the effect of dynamical friction which regulates the evolution of the bar and is mainly triggered by the live DM halo that environs the bar is not present in modified gravity models. This difference between modified gravity and live DM halo models is vividly visible in the time evolution of $\mathcal{R}$ parameter. As the main result of this paper, we confirm that the modified gravity models predict fast bars in contrast to the DM models. Of course, this means that modified gravity is in better agreement with observation. More importantly, this fact is independent of the initial disk's mass distribution.

Investigation of the vertical behavior of our models in the inner regions, reveals the existence of the buckling instability in all the models. The buckling instability is more efficient in the DM models in the sense that the final value of RMS thickness of the DM models is higher than NLG and MOG at the end of the simulation. It is necessary to reiterate that this difference is also reported in \cite{2018ApJ...854...38R} and \cite{2019ApJ...872....6R} for exponential disks. On the other hand, studying the vertical behavior of the models in the outer regions illustrates that this parameter has higher values in our modified gravity models in comparison to the DM models. As presented in the edge-on views, the modified gravity models are more warped and flared which is compatible with the previous results presented in \cite{2007A&A...464..517T}. 

According to the measured parameters, it could be concluded that apart from the early stages of the disk evolution, the general behavior of the bar instability, pattern speed, and also vertical structure, does not change significantly by changing the disk surface density in modified gravity models. Furthermore, the substantial differences with disks surrounded by DM, remain unchanged. In other words, we emphasize that in analogy to the results presented in \cite{2018ApJ...854...38R} and \cite{2019ApJ...872....6R}, the difference in the evolution of the disk galaxies under the effect of modified gravity and live DM halo is independent of the employed density profile of the baryonic disk and DM halo.    

\section*{Acknowledgement}
We would like to thank the unknown referee for his/her instructive comments which helped us a lot to improve this work. NG would like to thank Victor Debattista for suggesting a method for calculating the bar length, which is essential in the study of $\mathcal{R}$ parameter. MR would like to appreciate  B. Mashhoon for stimulating discussions on nonlocal gravity. We would also like to thank Benoit Famaey for the suggestion of a helpful reference. This work is supported by Ferdowsi University of Mashhad under grant No. 1/50936 (14/08/1398).

\bibliographystyle{apj}
\bibliography{short,galaxycheck_rev_2}
\label{lastpage}
\end{document}